\documentclass[11pt]{article}

\pdfoutput=1 
\usepackage{./aux/jheppub}
\usepackage[utf8]{inputenc}
\usepackage{graphicx,tikz}
\usepackage{amsfonts,amsmath,amssymb}
\usepackage{array,mathrsfs,yfonts,dsfont,bbm,colonequals,amscd,mathtools}
\usepackage{relsize,suffix,cancel,bbm,capt-of,upgreek,verbatim}

\newcommand{\be}{\begin{equation}}
\newcommand{\ee}{\end{equation}}
\newcommand{\ba}{\begin{equation}\begin{aligned}}
\newcommand{\ea}{\end{aligned}\end{equation}}

\usepackage{mathtools}          

\DeclareMathOperator*{\Res}{Res}
\DeclareMathOperator*{\vol}{vol}

\usepackage{amssymb}  
\colorlet{darkblue}{blue!70!black}
\colorlet{darkgreen}{green!70!black}
\colorlet{darkred}{red!70!black}

\title{Sharp Boundaries for the Swampland}

\preprint{
YITP-SB-21-3\\
\vspace{-0.34in}
\begin{flushright}CALT-TH 2021-009\end{flushright}
}

\author[a]{Simon Caron-Huot,\!}
\author[b]{Dalimil Maz\'{a}\v{c},\!}
\author[c]{Leonardo Rastelli,\!}
\author[d]{and David Simmons-Duffin}

\affiliation[a]{Department of Physics, McGill University, 3600 Rue University, Montr\'eal, QC Canada}
\affiliation[b]{Institute for Advanced Study, Princeton, NJ 08540, USA}
\affiliation[c]{C. N. Yang Institute for Theoretical Physics, Stony Brook University,
Stony Brook, NY 11794, USA}
\affiliation[d]{Walter Burke Institute for Theoretical Physics, Caltech, Pasadena, CA 91125, USA }

\abstract{
We reconsider the problem of bounding higher derivative couplings in consistent weakly coupled gravitational theories, starting from general assumptions about analyticity and Regge growth of the S-matrix. 
Higher derivative couplings are expected to be of order one in the units of the UV cutoff. Our approach justifies this expectation and allows to prove precise bounds on the order one coefficients. Our main tool are dispersive sum rules for the S-matrix. We overcome the difficulties presented by the graviton pole by measuring couplings at small impact parameter, rather than in the forward limit. We illustrate the method in theories containing a massless scalar coupled to gravity, and in theories with maximal supersymmetry.
}


\usepackage[T1]{fontenc}
\usepackage{dsdshorthand}

\begin{document} 

\maketitle
\flushbottom



\section{Introduction}\label{sec:Introduction}

The gravitational coupling grows at short distances, giving rise to the fundamental question: how is gravity UV completed? In the absence of experiments directly probing the Planck scale, a  more relevant
question is: what is the space of low-energy effective field theories (EFTs) that contain gravitons and admit a UV completion?   There are compelling reasons to believe that 
the rules of the game are more stringent in the presence of dynamical gravity.  A variety of conceptual arguments, reinforced by surveys of the string theory landscape, have led to  intriguing ``swampland'' conjectures~\cite{Ooguri:2006in, Brennan:2017rbf, Palti:2019pca, vanBeest:2021lhn} (such as the weak-gravity conjecture~\cite{ArkaniHamed:2006dz}),  which are proposed criteria for an EFT to be embedabble in a consistent theory of quantum gravity.  

Even {\it without} dynamical gravity, it has long been recognized that  ``not anything goes'' in effective field theory. In a local QFT in flat space, the  constraints of unitarity and causality imply   inequalities for the low-energy effective couplings~\cite{Pham:1985cr,Adams:2006sv}. 
A series of recent papers~\cite{Arkani-Hamed:2020blm,Bellazzini:2020cot, Tolley:2020gtv, Caron-Huot:2020cmc,Sinha:2020win} initiated a  systematic analysis of the inequalities that follow from consideration  of $2 \to 2$ scattering processes. The analysis begins with canonical assumptions about the S-matrix, such as
analyticity, crossing and, crucially,  Regge boundedness. For a fixed momentum transfer $u < 0$, the amplitude ${\cal M}(s, u)$ is assumed to grow slower than $|s|^2$ at large $|s|$ in the upper-half  complex plane.\footnote{
This behavior is implied by the stronger Froissart-Martin  bound~\cite{Froissart:1961ux, Martin:1962rt}, ${\cal M}(s, u)  \lesssim
 s \log^{D-2} s$, which (at least in a gapped theory) is a consequence of unitarity if one assumes  polynomial boundedness and analyticity, which in turn can be established from first principles in a local QFT.  See Appendix A.2 of \cite{Arkani-Hamed:2020blm}  for a nice recent discussion.} This is sufficient to derive twice-subtracted dispersion relations \cite{PhysRev.95.1612,PhysRev.99.979}. These can be in turn used to express the parameters of the EFT (valid at energies $E \ll M$, where $M$ is the UV cut-off) in terms of the UV data that enter for $E \geq M$. One is agnostic about the UV physics,  except that it contributes to the partial wave expansion of ${\cal M} (s, u)$  with a definite sign, thanks to unitarity. 
    These positive sum rules constrain the allowed low-energy couplings. Notably, one can  establish two-sided bounds~\cite{Tolley:2020gtv, Caron-Huot:2020cmc} that put dimensional analysis on a firm footing. A low-energy parameter
    of mass dimension $-\alpha$ must scale like $\sim 1/M^\alpha$, possibly further suppressed by a small coupling but never larger,  and with a rigorous estimate of the $O(1)$ numerical factor.
        
It is very natural to ask whether the same approach may work in EFTs containing gravitons, i.e. massless spin-two particles. Can one develop a systematic and quantitative approach to the swampland,  relying only on general properties of the S-matrix in asymptotically flat spacetime?  In the presence of dynamical gravity,  the analyticity and boundedness properties of the S-matrix are admittedly more speculative, and we will treat them as postulates. We will in particular assume that the same Regge bound $\lim_{ s \to \infty} {\cal M}(s, u) / s^2 = 0$  (for fixed $u <0$) holds also in this case. Note that this is a little stronger than the $O(s^2)$ bound assumed in the ``Classical Regge Growth'' conjecture~\cite{Camanho:2014apa, Chowdhury:2019kaq, Chandorkar:2021viw}, which is believed to hold for any consistent {\it tree-level} S-matrix; we need to require that at the nonperturbative level the Regge growth is {strictly} smaller than $s^2$. The best heuristic justification for such a behavior comes from physical arguments
that the scattering amplitude in impact parameter space should be analytic and bounded,
as a consequence of unitarity and causality. The current best technical justification comes from viewing the S-matrix in  asymptotic $D$-dimensional Minkowski space as the flat-space limit of a scattering process in asymptotic $\text{AdS}_D$ space, according to the prescription of \cite{Susskind:1998vk, Polchinski:1999ry, Penedones:2010ue}.
 On the AdS side, we can appeal to the rigorous  
 non-perturbative bound on the  Regge behavior of the dual CFT correlator~\cite{Caron-Huot:2017vep}, which when translated to  flat-space variables corresponds to an $O(s)$ bound.\footnote{At leading order for large $N$, the bound on the Regge
 behavior of the CFT correlator is known as the chaos bound~\cite{Maldacena:2015waa}. The chaos bound translates precisely  to the $O(s^2)$ bound of the Classical Regge Growth conjecture, as has been carefully argued in the recent paper \cite{Chandorkar:2021viw}. It would be very interesting to see whether the arguments of \cite{Chandorkar:2021viw} can be extended to the nonperturbative regime and give a rigorous justification for our assumption that the Regge growth is {\it strictly better} than $O(s^2)$.}

As a proof of principle, we focus on the simplest example:  $2 \to 2$ scattering of identical massless scalars coupled to gravity. We assume that the low-energy EFT is weakly coupled, but make no assumptions about the UV physics beyond analyticity and unitarity. An important example to which our analysis applies is string theory with fixed but small coupling $g_s \ll 1$. The cutoff $M$ is the string scale, i.e.~the mass of the first exchanged massive string mode. The theory is weakly coupled at and below the cutoff, but it eventually becomes strongly coupled as we approach the Planck scale $M/ (g_s)^{\frac{2}{D-2}}$. 

It is straightforward to derive dispersive sum rules in this setup, following the blueprint reviewed in \cite{Caron-Huot:2020cmc}. Similar positive sum rules incorporating gravity were written down before, see e.g.~\cite{Bellazzini:2019xts, Tokuda:2020mlf, Arkani-Hamed:2020blm,Alberte:2020jsk,Pajer:2020wnj,Grall:2021xxm} (or \cite{Alberte:2020bdz} for the similar case of $U(1)$ gauge theory).
There is however a notorious obstacle in deriving bounds that involve the Newton constant $G$: the
graviton propagator diverges in the limit of forward scattering ($u \to 0$ in our conventions), seemingly invalidating the application of the positivity constraints. We circumvent this problem by measuring couplings at small impact parameter 
$b \ll 1/M$, while keeping the momentum transfer $u \sim - M^2$. The physical picture of a scattering experiment at small impact parameter is the same as in \cite{Camanho:2014apa}, but now in the more systematic framework of the S-matrix bootstrap where we look for numerical bounds (not only parametric).
Apart from curing the graviton divergence, this approach has  the virtue of making dimensional analysis transparent. Finally, while in this paper we treat the EFT at tree level, it is in principle straightforward to include EFT loop corrections, which will introduce low-energy cuts extending all the way to $u=0$. Our method remains valid even if the amplitude is non-analytic at $u = 0$ and it is thus ideally suited to handle loops.
 
As in previous work, the positivity constraints lead to a linear programming problem that can be implemented numerically to carve out the space of EFT couplings. See Figure~\ref{fig:allowedregionsdifferentD} and 
 Figure~\ref{fig:highercontacts} for some sample exclusion plots for   ratios of the scalar higher derivative EFT coefficients $g_k$ over the Newton constant $G$, in units of the UV cutoff. We restrict our analysis to spacetime dimension $D > 4$, because for $D = 4$ the integral transform to impact parameter space suffers from an IR divergence arising from the massless graviton pole. We believe that it should be possible to 
 extend our approach to $D=4$,  working with suitable IR-safe observables.  
 
We also consider bounds on graviton scattering in the presence of maximal supersymmetry. Thanks to supersymmetry, $2 \to 2$ graviton scattering can be expressed in terms of an auxiliary fully crossing symmetric scalar amplitude with improved Regge behavior. We derive a numerical upper bound on the coefficient of the  leading $R^4$ curvature correction, and check  that it is obeyed by type II string theory.

In an upcoming paper \cite{CHMRSD3}, we will study the analogous problem of bounding higher derivative couplings in a gravitational theory in AdS space. A prominent physical question~\cite{Heemskerk:2009pn} is to show
that a large $N$ CFT with a large gap admits a local bulk dual, i.e. such that higher-derivative corrections are  suppressed by inverse powers of the gap.
The formalism of dispersive CFT sum rules \cite{Carmi:2019cub, Mazac:2019shk, Penedones:2019tng, Caron-Huot:2020adz,Gopakumar:2021dvg}\footnote{See also \cite{Mazac:2016qev,Mazac:2018mdx,Mazac:2018ycv,Paulos:2019gtx,Carmi:2020ekr} for more work concerning dispersive sum rules in CFTs.} allows for an almost direct uplift of our flat space results to AdS. While the logic  is very similar,  the  details of the AdS story are technically more involved. The flat space analysis presented here will serve as an indispensable  warmup. One distinct advantage of the AdS setup is that all the requisite analyticity and boundedness properties are rigorous consequences of the CFT axioms.

\bigskip

The rest of the paper is organized as follows. In Section \ref{sec:flatspace}, we state our assumptions and review the approach of \cite{Caron-Huot:2020cmc} to bounding EFT coefficients, which relies on the Taylor expansion of the dispersive sum rules around the forward limit. In Section \ref{sec:secondstrategy}, we develop our new strategy, based on localization at small impact parameter, and apply it to simple examples of gravitational EFTs. We conclude in Section \ref{sec:conclusions}. In Appendix~\ref{app:flatspacenumerics}, we present some technical details of the numerical implementation of the linear program. Finally, in  Appendix~\ref{app:extendedu} we consider bounds coming from extending the range of the momentum transfer beyond $|u|=M^2$.


\def\GG{\mathcal{G}}
\def\MM{\mathcal{M}}
\def\hatPi{\widehat{\Pi}}
\def\legP{\mathcal{P}}
\def\JJ{\mathcal{J}^2}
\def\JJJ{\mathcal{J}^4}
\def\avg#1{\Big\langle#1\Big\rangle}
\newcommand\SCH[1]{{\color{blue} \it SCH: #1}}

\section{Dispersive sum rules}\label{sec:flatspace}

In this section we state our assumptions and set the stage for the main argument in Section~\ref{sec:secondstrategy}. We will also review how to bound couplings in non-gravitational theories using dispersion relations expanded around the forward limit. We will closely follow the presentation of \cite{Caron-Huot:2020cmc}.

\subsection{Assumptions and first consequences}
\label{ssec: review disp sum rules}

\medskip
\noindent
{\it  IR Effective Field Theory}
\medskip

\noindent 
Consider a real scalar coupled to gravity and to unknown heavy states of mass greater than some energy scale $M$, in asymptotically flat $D$-dimensional spacetime.
As our interest lies in the effects of the heavy scale $M$,  we take the scalar to be massless, though our techniques are general and can be applied equally well to massive scalars. The 
2 $\to$ 2 massless scalar scattering amplitude takes the form
\be\begin{aligned} \label{Mlow param}
\MM_{\rm low}(s,u) &=\ 8\pi G \left[ \frac{st}{u} + \frac{su}{t} + \frac{tu}{s}\right] - \lambda_3^2\left[ \frac{1}{s}+\frac{1}{t}+\frac{1}{u}\right]
-\lambda_4 \\& \quad + g_2 (s^2+t^2+u^2) + g_3 (stu) + g_4 (s^2+t^2+u^2)^2+\ldots \, ,
\end{aligned}\ee
where
\be
s= - (p_1+p_2)^2\,,\quad t=- (p_1-p_4)^2\,,\quad u=- (p_1-p_3)^2\, 
\ee
are Mandelstam invariants, satisfying $s + t + u = 0$. $G$ is Newton's constant and the first term represents the exchange of gravitons in the three channels. Apart from coupling to gravity, we have also allowed for the possibility of scalar self-interactions. 

\begin{figure}[h]
\centering
\includegraphics[width=0.75\textwidth]{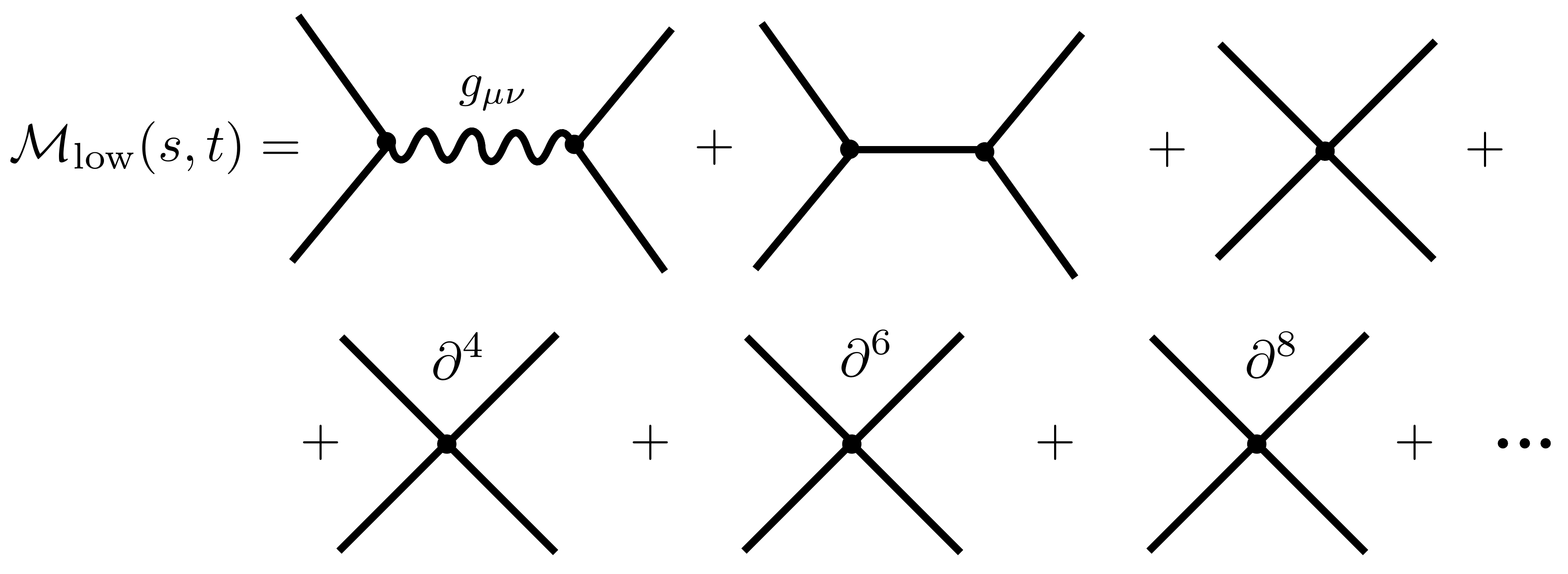}
\caption{A schematic representation of the low-energy amplitude \eqref{Mlow param}. The external scalar particles interact through graviton exchange, scalar exchange and a series of higher-derivative contact interactions.}
\label{fig:Scattering}
\end{figure}

We assume that the low-energy physics is weakly coupled. To make this precise, we can assume that there is a family of S-matrices parametrized by a coupling $\epsilon>0$ such that all the interactions, such as $G, \,\lambda_3^2, \,g_2$ etc. in \eqref{Mlow param}, are $O(\epsilon)$ as $\epsilon\rightarrow0$. Eq.~\eqref{Mlow param} is then the low-energy form of the scattering amplitude at the leading order in $\epsilon$, i.e.\ at tree level. For example, this is the situation in string theory, the role of $\epsilon$ being played by the string coupling. Low-energy loop corrections are $O(\epsilon^2)$ effects and could be included in principle but will be beside our main concern. In the rest of this paper, we will work at the leading order in $\epsilon$ and therefore these corrections will be absent.

Under this weak-coupling assumption, the 
EFT coefficients are symmetrical polynomials in the Mandelstam variables, corresponding to crossing-symmetric contact diagrams.
We have chosen the natural basis of monomials $\MM_{a,b}(s,t)=A^a B^b$,
where $A= s^2+t^2+u^2$, $B=s t u$. The spin of the monomial $\MM_{a,b}(s,t)$ is $J = 2a+2b$ and the scaling dimension is $4a+6b$.\footnote{The spin of a contact diagram is defined as the maximal spin appearing in its partial wave decomposition. We define the scaling dimension as the number of derivatives in the corresponding term in the Lagrangian. This differs from the standard definition by a constant shift.} Thus, there are $(J+2)/2$ independent contacts with spin equal to $J$, namely
\be
A^{\frac{J}{2}}B^0\,,\quad A^{\frac{J}{2}-1}B^1\,,\quad\ldots\,,\quad A^{0}B^{\frac{J}{2}}\, ,
\ee
with scaling dimensions
\be
2J\,,\quad 2J+2\,,\quad\ldots\,,\quad 3J\,.
\ee
Note that there can be multiple contacts with different spins but the same scaling dimension, the lowest example being $A^3$ and $B^2$.  The subscript $k$ in the EFT coefficients $g_k$  indicates (half) their scaling dimension, 
which specifies the contacts diagrams unambiguosuly for $k \leq 5$; for 
higher value of $k$
we  introduce additional labels (e.g.,~$g_6$ and $g'_6$) to distinguish contacts with degenerate scaling dimension.

\medskip
\noindent
{\it  UV unitarity constraints}
\medskip

\noindent 
Assuming there are no new degrees of freedom below the heavy scale $M$,
we want to know: what are the implications of high-energy unitarity on the coefficients $g_{k}$? It is natural to consider ratios of couplings such as
\be
\frac{g_n}{g_2}\quad\text{or}\quad\frac{g_n}{G}\,,
\ee
which stay finite as $\epsilon\rightarrow 0$. Naive EFT scaling, or dimensional analysis, suggests that 
\be \frac{g_{k}}{g_2} =  \frac{c_{k}}{M^{2(k-2)}}
\ee
where $c_{k}$ are dimensionless coefficients of order one. 
This is certainly the case if the EFT is obtained by integrating out a single heavy particle of mass $M$. However, we will be agnostic about the detailed physics above the heavy scale $M$, and only assume that the amplitude is unitarity and causal. In particular, it will {\it not\/} be important to assume that the heavy physics is weakly coupled at arbitrarily high energy scales. On the other hand, we will need \eqref{Mlow param} to be valid up to energies of order $M$, which does require the physics there to be weakly coupled. For example, in string theory with small but finite $g_s$, the appropriate cutoff $M$ is the string scale, where the theory is still weakly coupled. (However, the theory  does become strongly-coupled at the Planck scale.)

Unitarity is simplest to state in a decomposition of the amplitude in angular momentum partial waves.  In the s-channel physical region $\{ s > 0 \, , -s < u < 0 \}$,
Mandelstam  $s$ gives the squared center-of mass energy, 
 while the scattering angle is
\be
\cos \theta = 1 +  \frac{2 u}{s} \,.
\ee
The partial wave decomposition reads
\be \MM(s,u) =
s^{\frac{4-D}{2}} \sum_{J\ {\rm even}} n^{(D)}_J c_J (s)  \legP_J\p{1+\frac{2u}{s}}\,,
\ee
where
$\legP_J(x)$ are proportional to Gegenbauer polynomials (and reduce to Legendre polynomials for $D=4$),
\be \legP_J(x) \equiv {}_2F_1\p{-J,J+D-3,\tfrac{D-2}{2},\tfrac{1-x}{2}}\ .\ee
The normalization
\be
n_J^{(D)}\equiv \frac{2^{D} \pi^{\frac{D-2}{2}}}{\Gamma(\tfrac{D-2}{2})} (J+1)_{D-4} (2J+D-3)
\ee
has been chosen (see e.g.~\cite{Giddings:2007qq,Correia:2020xtr})  such that  unitarity of the S-matrix,
\be
 S S^\dagger = 1\, , \quad S = 1 + i \MM \, ,
 \ee
translates into $ |1+i c_J (s) |^2\leq 1$. Defining the spectral density $\rho_J (s) = {\rm Im}\, c_J(s)$, we can also write
\be \label{ImMpartial}
{\rm Im} \,  \MM(s,u)  = s^{\frac{4-D}{2}} \sum_{J\ {\rm even}} n^{(D)}_J \rho_J(s)\, \legP_J\p{1+\frac{2u}{s}} \, ,
\ee
where the unitarity constraint reads
\be
0 \leq \rho_J(s) \leq 2 \, , \qquad s > 0 \,, \;J \;{\rm even}\,.
\ee
The crucial fact is that ${\rm Im}\ \MM(s,u)$ is a positive sum of Gegenbauer polynomials.

We can now give a precise definition of the scale $M$. It is the energy where $\MM(s,u)$ first develops a nonzero imaginary part. In other words, we will assume that $\rho_J(s)$ vanishes for all $0<s<M^2$ and all even $J$.\footnote{Note that this definition only makes sense at the leading order at weak coupling since EFT loops give rise to an imaginary part for any $s>0$. However, as explained above, this is a subleading effect under our assumptions.}

\medskip
\noindent
{\it  Dispersive sum rules}
\medskip

\noindent 
A link between the regimes of high- and low-energy is provided by a dispersion relation. We assume that for fixed real $u < 0$, 
\begin{enumerate}
\item[(i)]  the amplitude $\MM(s, u)$ is analytic in $s$   in the upper-half plane ${\rm Im} \,s > 0$,\footnote{The amplitude is extended to the  lower-half complex $s$ plane 
 by ${\cal M} (s^* , u^*) =  {\cal M}^* (s , u)$, so for fixed $u <0$ it is analytic in $s$  away from the real axis.} and
\item[(ii)]  
 $\MM(s, u)$ has spin-2 convergence
in the  Regge limit, meaning that 
 the following limit vanishes
along any line of constant phase,
\be
 \lim_{|s|\to\infty} \frac{\MM(s,u)}{s^2} = 0\qquad (u<0)\,.
 \label{eq:ReggeBound}
\ee
\end{enumerate}
For example, in string theory, (\ref{eq:ReggeBound}) is ensured by Reggeization of the graviton: $\MM=O(s^{2+\alpha' u})$. We have discussed this crucial assumption in the Introduction.

\begin{figure}
\centering
\includegraphics[width=\textwidth]{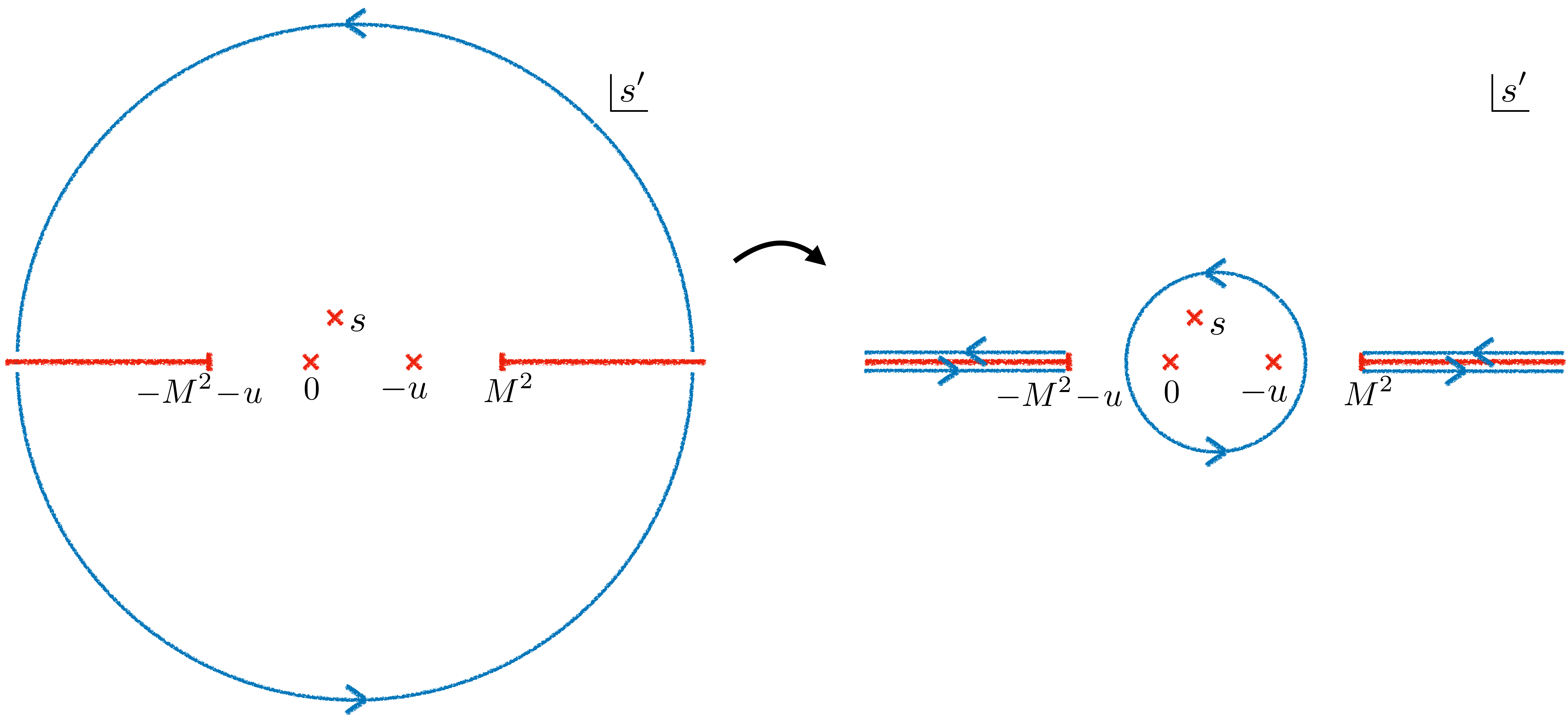}
\caption{Contour deformation leading to the dispersive sum rule \eqref{eq:DispersiveSumRule1}. We start from an integral over a large circle \eqref{starting}, which vanishes due to the spin-2 boundedness assumption \eqref{eq:ReggeBound}. After the contour deformation, we end up with an IR contribution, represented by the small circle on the right, and a UV contribution, represented by an integral over the cuts starting at $s'=M^2,-M^2-u$. In the presence of EFT loops, the IR contribution would also include cuts stretching between $s'=-M^2-u$ and $s'=M^2$, but these are subleading under our assumptions. Since the theory is assumed weakly coupled at scale $M$, the heavy cut is well approximated by a discrete set of poles for $s'$ not much larger than $M^2$. However, this will play no role in our reasoning.}
\label{fig:DispersionContour}
\end{figure}

Together these conditions imply the existence of twice-subtracted dispersion relations. 
The starting point is
\be \label{starting}
 \oint_{\infty} \frac{ds'}{2\pi i(s'-s)} \frac{\MM(s',u)}{s'(s' + u)} = 0 \, ,
\ee
where the contour integral is over a large circle. We picked $s'=0$ and $s' = - u$ as subtraction points -- this is a natural choice because it maintains  the $s \leftrightarrow t$ crossing symmetry without introducing an extraneous mass scale into the problem. We now deform the contour as in Figure \ref{fig:DispersionContour}. We take $|s|\ll M^2$. There are two kinds of contributions:
a low-energy circle at $|s'|\sim M^2$ which encloses the residues at $s'=s$, $s'=0$ and $s'=-u$;
and the contributions from the s-channel and t-channel high-energy cuts, starting respectively at $s' = M^2$ and $s'=  -M^2 - u$. Separating the low- and high-energy contributions,
\ba
& \frac{\MM_{\rm low}(s,u)}{s(s+u)} + \Res\limits_{s'=0} \left[ \left(\frac{1}{s'-s} + \frac{1}{s'+s+u}\right) \frac{\MM_{\rm low}(s',u)}{s'(s'+u)}\right]=\\
  & = \int_{M^2}^\infty \frac{ds'}{\pi} \left(\frac{1}{s'-s} + \frac{1}{s'+s+u}\right) \ {\rm Im}\left[\frac{\MM_{\rm high}(s',u)}{s'(s'+u)}\right]\,,
\label{eq:DispersiveSumRule1}
\ea
where we have used the $s \leftrightarrow t$  symmetry to combine the contributions of the two low-energy residues at $s'=0$ and $s'= -u$,
and the contributions of the two  high-energy cuts. 
For the terms on the left-hand side we use the low-energy parametrization (\ref{Mlow param}),
while for the heavy contribution, about whose details we are agnostic, we insert the partial wave decomposition (\ref{ImMpartial}),
\be 
\frac{8\pi G}{-u} +2g_2 -g_3 u+4g_4(2 u^2+s (s+u))+\ldots
= 
\avg{
\frac{(2m^2+u)\ \legP_J(1+\frac{2u}{m^2})}{(m^2+u)(m^2-s)(m^2+s+u)} }\,, \label{B2 flat A}
\ee
where we defined the heavy averages as 
\be
 \avg{(\cdots)} \equiv \frac{1}{\pi} \sum_{J\, {\rm even}}  n^{(D)}_J
 \int_{M^2}^\infty \frac{dm^2}{m^2} m^{4-D}  \rho_J(m^2) \;(\cdots)\,.
 \label{eq:UVaverage}
\ee
What will be important for us is that this is a positive measure since $ \rho_J (m^2) \geq 0$ by unitarity. We make no use of the upper bound $ \rho_J (m^2) \leq 2$ in this work.
Note that the $s$- and $t-$ channel poles have cancelled in \eqref{B2 flat A},
leaving a regular function that can be Taylor-expanded
in powers of $s/M^2$. 

We will organize these sum rules by expanding around small $s$.
Because our subtraction preserves $s\leftrightarrow t$ symmetry, it is easy to see that the series proceeds in integer powers of $s(s+u)$ and that we can expand (\ref{B2 flat A})
as 
\be
\sum_{n=1}^\infty [ s (s+u)]^{n-1} {\cal C}_{2n, u} = 0 \,.
\ee
For example, the first sum rule, denoted by ${\cal C}_{2, u}$, is obtained by taking the $s \to 0$ limit in~(\ref{B2 flat A}),
\be
{\cal C}_{2, u} = \frac{8\pi G}{-u} +2g_2 -g_3 u+8g_4 u^2+\ldots -
\avg{\frac{(2m^2+u)\legP_J(1+\frac{2u}{m^2})}{m^2(m^2+u)^2}}  \label{B2 flat}\, = 0\, ,
\ee
while the next sum rule ${\cal C}_{4, u}$ corresponds to the coefficient of $s(s+u)$, 
\be
{\cal C}_{4, u} =  4g_4 +\ldots - \avg{\frac{(2m^2+u)\legP_J(1+\frac{2u}{m^2})}{m^4(m^2+u)^3}}=0 \, .
\label{B4 flat}
\ee
We can also obtain the ${\cal C}_{k, u}$ sum rules more directly from the spin-$k$ subtracted version of the  dispersion integral,\footnote{Compare with the $s\to 0$ limit of (\ref{starting}).}
\be \label{Ckudirect}
{\cal C}_{k, u} \equiv  \oint_{\infty} \frac{ds'}{2\pi i}  \frac{1}{s'} \frac{\MM(s',u)}{[s'(s' + u)]^{k/2}} = 0 \, ,\qquad u<0,\ k =2, 4, \dots \,.
\ee
By  the same contour deformation argument  as above, we find the explicit expressions 
\be
\label{Bks}
{\cal C}_{k, u} = \Res\limits_{s'=0} \left[\frac{2s' + u}{s'(s'+u)}\frac{\MM_\text{low}(s',u)}{\big[s'(s'+u)\big]^{k/2}}\right]
	- \avg{ \frac{2m^2 + u}{m^2+u} \frac{\legP_J \left(1+\tfrac{2 u}{m^2}\right)}{\big[m^2(m^2+u)\big]^{k/2}}}\, .
	\ee
 The ${\cal C}_{k, u}$ sum rule only receives contributions from the contacts 
 \be
 (s^2 + t^2 + u^2)^a (stu)^b \,  \quad {\rm with} \;{\rm spin}\; J = 2a + 2b \geq k \; {\rm and}\; b \leq k/2 \, .
 \ee 
The condition $J \geq k$ is seen by closing the contour around infinity, as in  (\ref{Ckudirect}), while the condition $b \leq k/2$ is apparent by closing the contour around the origin, as in (\ref{Bks}).
Finally, note that only the ${\cal C}_{2, u}$ sum rule  is sensitive to the graviton exchange, through the term  $-8 \pi G /u$.

We now describe two distinct ways to use the ${\cal C}_{k,u}$ sum rules. First, in Section~\ref{ssec:forwardLimit} we review bounds obtained by expanding around the forward limit. Then, in Section~\ref{sec:secondstrategy} we introduce a new method: localizing in impact parameter space.

\subsection{Review: Bounds from the forward limit}
\label{ssec:forwardLimit}

We now briefly illustrate the strategy described in \cite{Tolley:2020gtv,Caron-Huot:2020cmc}
to derive inequalities for the EFT coefficients by expanding the sum rules (\ref{Bks}) in the forward limit $u  \to 0$. This strategy breaks down in the presence of gravity, because the ${\cal C}_2 (u)$ sum rule diverges in the forward limit, and so in this subsection we switch off gravity by setting $G=0$.

Derivatives of ${\cal C}_{k, u}$ with respect to $u$ at $u=0$ compute the couplings $g_2$, $g_3$, \ldots. Only couplings of spin two and higher appear in the sum rules. Spin-two couplings $g_2$ and $g_3$ only appear in ${\cal C}_{2, u}$ and they are computed respectively by $\cC_{2,0}$ and $\cC'_{2,0}$\footnote{Here and below, primes denote derivatives with respect to $u$.}
\be
g_2 = \left\langle\frac{1}{m^4}\right\rangle\,,\qquad
g_3 = \left\langle\frac{3-\frac{4\mathcal{J}^2}{D-2}}{m^6}\right\rangle\,,
\label{eq:g2g3}
\ee
where $\mathcal{J}^2 = J(J+D-3)$ is the quadratic Casimir of the massive little group $SO(D-1)$. The spin-4 couling $g_4$ appears in $\cC_{2,u}$ and $\cC_{4,u}$ and is computed by $\cC''_{2,0}$ and $\cC_{4,0}$
\be
g_4 = \left\langle\frac{1+\frac{\mathcal{J}^2(2\mathcal{J}^2-5D+4)}{2D(D-2)}}{2m^8}\right\rangle = 
\left\langle\frac{1}{2m^8}\right\rangle\,.
\ee
The last equality leads to the simplest example of a null constraint on the heavy data
\be
\left\langle\frac{\mathcal{J}^2(2\mathcal{J}^2-5D+4)}{m^8}\right\rangle = 0\,.
\ee
Null constraints arise because the low-energy amplitude is symmetric under $s\leftrightarrow u$.

The first sum rule in \eqref{eq:g2g3} immediately implies $g_2>0$. The remaining sum rules can be used to derive two-sided bounds on the ratios $g_n/g_2$. It is convenient to normalize the measure \eqref{eq:UVaverage} by the $g_2$ sum rule and define the probablity measure $\tilde{\rho}_{J}(m)$
\ba
 \avg{(\cdots)}_{g_2} &=\frac{1}{\pi g_2}\sum_{J\, {\rm even}}  n^{(D)}_J
 \int_{M^2}^\infty \frac{dm^2}{m^2} m^{-D}  \rho_J(m^2) \;(\cdots) \\
 &\equiv \sum\limits_{J\text{ even}}\,\int\limits_{M}^{\infty}\!dm \,\tilde{\rho}_{J}(m)\;(\cdots)
 \,.
 \label{eq:g2average}
\ea
The above sum rules become
\be
1 = \langle 1 \rangle_{g_2}\qquad
\frac{g_3}{g_2} = \left\langle \frac{3-\frac{4\mathcal{J}^2}{D-2}}{m^2} \right\rangle_{g_2}
\label{eq:g2g3Normalized}
\ee
and
\be
\frac{g_4}{g_2} = \left\langle \frac{1}{2m^4} \right\rangle_{g_2}\qquad
0 = \left\langle \frac{\mathcal{J}^2(2\mathcal{J}^2-5D+4)}{m^4} \right\rangle_{g_2}
\ee
Since the measure is supported in $m\geq M$, it immediately follows from the second equation in \eqref{eq:g2g3Normalized} that
\be
\frac{g_3}{g_2}\leq \frac{3}{M^2}\,.
\ee

One can derive a lower bound on $g_3/g_2$ with the help of the null constraint. The $g_2$, $g_3$ and null sum rules take the form of a vector equation
\be
\sum\limits_{J\text{ even}}\,\int\limits_{M}^{\infty}\!dm \,\tilde{\rho}_{J}(m)
\begin{pmatrix}
1\\
\frac{1}{m^2}\left(3-\frac{4\mathcal{J}^2}{D-2}\right)\\
\frac{\mathcal{J}^2(2\mathcal{J}^2-5D+4)}{m^4}
\end{pmatrix}
=
\begin{pmatrix}
1 \\ g_3/g_2 \\ 0
\end{pmatrix}\,.
\ee
This is now a standard linear programming problem. We are asking for what values of $g_3/g_2$ is the vector on the RHS in the positive cone spanned by the vectors on the LHS. This means that the allowed range for $g_3/g_2$ is the intersection of the convex hull of points
\be
\begin{pmatrix}
\frac{1}{m^2}\left(3-\frac{4\mathcal{J}^2}{D-2}\right)\\
\frac{\mathcal{J}^2(2\mathcal{J}^2-5D+4)}{m^4}
\end{pmatrix} \in \mathbb{R}^2
\ee
with the $x$-axis. It is clear that the region is finite, and easy to check that the lower bound comes from considering only $J=2$ and $J=4$ trajectories. In other words, including $J>4$ trajectories does not increase the size of the intersection of the convex hull with the $x$-axis. The bound thus comes from the intersection of the $x$-axis and the line connecting the point $J=2$, $m=M$ with the point $J=4$, $m=m_4$ and optimizing over $m_4$. It is clear that since the bound comes from bounded spin, it needs to have the bulk-point scaling $M^{-2}$. In fact, if there is any bound at all, it must have this scaling simply by dimensional analysis. Therefore one can get an analytic, if cumbersome answer:
\be
-\frac{\kappa(D)}{M^2}<\frac{g_3}{g_2}\leq\frac{3}{M^2}\,,
\ee
where
\be
\kappa(D) = \sqrt{\frac{(D+3) \left(319 D^3+76D^2- 292 D + 32\right)}{24(D-2)^2 (D+1) (D+4)}}+\frac{6 (5 D-2)}{12 (D-2)}\,.
\ee
The bound can be improved by considering combinations of more functionals, see \cite{Caron-Huot:2020cmc,Li:2021cjv}.

We can play the same game to bound $g_4/g_2$ but in this case including the null constraint buys us nothing since spin does not enter the sum rule for $g_4/g_2$. The result is simply
\be
0\leq \frac{g_4}{g_2} \leq \frac{1}{2M^4}\,.
\ee


\section{Bounds with gravity}
\label{sec:secondstrategy}

\subsection{General idea}
The preceding strategy, Taylor expanding around the forward limit, suffers from three drawbacks.
First, it does not manifest the expected scaling $g_3\sim \frac{g_2}{M^2}$ until the final stage.
This is because we are evaluating the IR contribution to dispersive sum rules at small momentum transfer $u$: $2g_2-3g_3 u +\ldots$. Second, Taylor series do not generalize naturally to handle loop corrections, which have branch cuts at $u=0$. Third, the strategy fails already at tree-level
in the presence of gravity, since the $\frac{8\pi G}{-u}$ pole in the ${\cal C}_{2,u}$ sum rule explodes in the forward limit.

We propose that these three issues admit a common physical resolution:
measure EFT couplings from small impact parameter scattering.
By doing measurements at impact parameter $b\sim 1/M$ and $u\sim -M^2$, the expected scaling will be
automatic, branch points are avoided, and the gravity pole will be suppressed.

Let us explain the mechanism for suppressing the gravity pole in more detail, and illustrate possible applications. The physical meaning of $u$ in dispersive sum rules $\cC_{k,u}$ is the magnitude of the momentum transfer. We have $u = -(p_3-p_1)^2$, and $\vec{p}= p_3-p_1$ is the spatial momentum transfer. For high-energy scattering, fixed impact parameter scattering, $\vec{p}=p_3-p_1$ lies in the $(D{-}2)$-dimensional plane transverse to the incident momenta. The impact parameter $\vec{b}$ is also a vector in $\mathbb{R}^{D-2}$. It is Fourier-conjugate to $\vec{p}$. We will write $p=|\vec{p}|$ and $b=|\vec{b}|$. We will use the following conventions for going between the momentum transfer and impact parameter space. Consider a spherically symmetric momentum-space wavefunction $f(p)$. We define the corresponding impact parameter space wavefunction as the $(D{-}2)$-dimensional Fourier transform
\be
\widehat f(b) \equiv 
\int d^{D-2}\vec p\, e^{i\vec b\.\vec p} \frac{f(p)}{ p^{D-3}\vol S^{D-3}} = \Gamma\p{\tfrac{D-2}{2}}  \int_0^\oo dp\,  f(p) \frac{J_{\frac{D-4}{2}}(p b)}{(pb/2)^{\frac{D-4}{2}}}.
\label{eq:impactparametertransform}
\ee
The factor $1/p^{D-3}$ was inserted for future convenience. An integral of $f(p)$ against the gravity pole $\frac{8\pi G}{-u}$ can be expressed in impact parameter space as
\be
\label{eq:impactparameterweightedintegral}
\int_0^\oo dp f(p) \frac{8\pi G}{p^2} = \int_0^\oo db \widehat f(b) \frac{8\pi G b}{D-4}\,.
\ee
Evaluation in the forward limit corresponds to $f(p)=\de(p)$, which in impact parameter space is $\widehat f(b)=1$. This leads to a divergence when integrated against the graviton contribution $\frac{8\pi G b}{D-4}$ in (\ref{eq:impactparameterweightedintegral}).  By contrast, if $\widehat f(b)$ is localized near $b\sim 1/M$, the contribution of the graviton pole will automatically be suppressed by $\sim 1/M^2$. An example wavefunction is shown in Figure~\ref{fig:wavepacket} below. Indeed, we will soon see that in dispersive bounds $\widehat f(b)$ is constrained to be positive, so localization in impact parameter space is the {\it only\/} way to suppress the graviton pole relative to other contributions. To achieve such localization, the momentum-space wavefunction must have support all the way up to $|u|\sim M^2$.

One might worry that there is a limitation on the range of $u$ due to the EFT series \eqref{B2 flat}
breaking down at $|u|\sim M^2$. Specifically, if we evaluate dispersion relations at $|u|\sim M^2$, all contact interactions could contribute equally and it would be difficult to disentangle individual EFT coefficients. This suggests restricting to $|u|\ll M^2$. But that would not be good enough to get numerical bounds with the right scaling in $M$ --- it would only give parametric bounds.

The key idea for getting around this difficulty is to use low-energy crossing symmetry to eliminate
all the terms starting from $g_4u^2$ and higher in the ${\cal C}_2$ sum rule. Note that the EFT contributions to the ${\cal C}_2,{\cal C}_4,{\cal C}_6$ sum rules are
\begin{align}
\left.{\cal C}_{2, u}\right|_\mathrm{EFT} &= \frac{8\pi G}{-u} +2 g_2-g_3u+8 g_4 u^2  -2g_5 u^3 + 24 g_6 u^4 -4 g_7 u^5 \dots\,,
\label{b2unimproved}\\
\left.{\cal C}_{4, u}\right|_\mathrm{EFT} &= 4 g_4-2 g_5 u +(24 g_6 + g_6')u^2  - 8 g_7 u^3+ \dots\,,
\\
\left.{\cal C}_{6, u}\right|_\mathrm{EFT} &= 8 g_6 -4 g_7 u + \dots\,.
\end{align}
By subtracting a linear combination of ${\cal C}_{4,0}$ and ${\cal C}_{4,0}'=\ptl_u {\cal C}_{4, u}|_{u=0}$ from ${\cal C}_{2, u}$, we can cancel the $g_4$ and $g_5$ terms in (\ref{b2unimproved}). Next, subtracting a linear combination of ${\cal C}_{6,0}$ and $ {\cal C}_{6,0}'$, we can cancel the $g_6$ and $g_7$ terms, and so on. Repeating this procedure, we find that the following linear combination of sum rules is independent of all higher EFT coefficients:
\begin{align}
\label{eq:definitionofb2improved}
{\cal C}_{2, u}^\mathrm{improved} &= {\cal C}_{2, u} - \sum_{n=2}^\oo \p{n\, u^{2n-2} {\cal C}_{2n,0} + u^{2n-1} {\cal C}_{2n,0}'}\,.
\end{align}
Specifically, we have
\begin{align}
\left.{\cal C}_{2, u}^\mathrm{improved}\right|_\mathrm{EFT} &= \frac{8\pi G}{-u} +2g_2 -g_3 u\,,
\end{align}
with no contamination by higher contact coefficients. 
Note that the improved sum rule  \eqref{eq:definitionofb2improved} still involves forward limits, but only of the higher-subtracted sum rules ${\cal C}_{4}, {\cal C}_{6},\ldots$, which do not have a graviton pole.  We suspect that it should be possible to find different improvements which eliminate forward limits altogether, but \eqref{eq:definitionofb2improved} will suffice for our purposes.
  
The contribution of heavy states to the ${\cal C}_{2, u}^\mathrm{improved}$ sum rule is found by inserting the heavy contribution from (\ref{Bks}) into (\ref{eq:definitionofb2improved}) and performing the sum over $n$.
This sum can be done in a closed form, yielding the following {\it exact} sum rule:
\ba
\frac{8\pi G}{-u} +2g_2 -g_3 u &=
\avg{\frac{(2m^2+u)\legP_J(1+\frac{2u}{m^2})}{m^2(m^2+u)^2}
- \frac{u^2}{m^6}\left(\frac{(4m^2+3u)\legP_J(1)}{(m^2+u)^2}+
\frac{4u \legP'_J(1)}{m^4-u^2}\right)}
\label{B2 improved flat} \\
&\equiv \avg{ {\cal C}_{2, u}^{\rm improved}[m^2,J] }\,.
\ea
The important feature of (\ref{B2 improved flat}) is that only three EFT couplings appear on the left, yet we retain
the full power of a one-parameter family of sum rules (labelled by $u$), which we can use to localize at small impact parameters. In the absence of gravity, $\mathcal{C}^\mathrm{improved}_{2,u}$ is equivalent to a combination of null constraints and evaluation around the forward limit
\begin{align}
\mathcal{C}^\mathrm{improved}_{2,u} &= \mathcal{C}_{2,0}+u \mathcal{C}_{2,0}' + u^2 \mathcal{X}_{2,u} \qquad \textrm{(without gravity)},
\end{align}
where $\mathcal{X}_{2,u}$ is written below in (\ref{eq:nullconstraints}).
However $\mathcal{C}^\mathrm{improved}_{2,u}$ makes sense even with gravity.

We expect that (\ref{B2 improved flat}), evaluated anywhere in the range $u\in (-M^2,0)$, gives a valid and convergent sum rule. When $-u=M^2$, the $t$-channel cut merges with the origin in the $s$-plane. Depending on the analytic structure of the $t$-channel cut, this may require us to modify the sum rule. Thus, we will mostly restrict to $u\in (-M^2,0]$. However, for meromorphic amplitudes, we expect that (\ref{B2 improved flat}) can have a larger range of validity. We discuss this idea further in Appendix~\ref{app:extendedu}.

We can now derive inequalities on the EFT couplings $8\pi G, g_2, g_3$ by constructing functions $f(p)$ whose integral against ${\cal C}^{\rm improved}_{2,-p^2}[m^2,J]$ is non-negative on all allowed heavy states:
\begin{align} \label{flat positivity}
\mbox{if}\quad& \int_0^M dp\, f(p) {\cal C}_{2,-p^2}^{\rm improved}[m^2,J] \geq 0 \qquad \forall m\geq M,\ J=0,2,4\ldots
\\
\mbox{then}\quad& \int_0^M dp\,f(p)\left[\frac{8\pi G}{p^2} +2g_2 +g_3p^2\right] \geq 0\,.
\label{eq:eftineq}
\end{align}
We will see that functions $f(p)$ allowed by the condition on the first line carve out an interesting region in the $(G,g_2,g_3)$ space. Let us interpret the first positivity condition. We claim that it \emph{implies} the statement that $f(p)$ is the Fourier transform of a positive function
of transverse impact parameter.  To see this, we take a scaling limit where $m\to\infty$ with the
``impact parameter'' $b=\frac{2J}{m}$ held fixed. In this limit, the Gegenbauer functions become Bessel functions as follows\footnote{This has the following simple interpretation. Gegenbauer functions are the basis for the harmonic decomposition of functions on the sphere $S^{D-2}$, while Bessel functions for the Fourier decomposition of radial functions on $\mathbb{R}^{D-2}$. \eqref{eq:GegenbauerToBessel} corresponds to the flat-space limit of the sphere with momentum in $\mathbb{R}^{D-2}$ fixed.}
\begin{align}
 \lim\limits_{m\to \infty} \legP_{\frac{mb}{2}}\p{1-\frac{2p^2}{m^2}} &=
 \frac{\Gamma\big(\tfrac{D-2}{2}\big)}{(bp/2)^{\frac{D-4}{2}}}J_{\frac{D-4}{2}}(bp)\,.
 \label{eq:GegenbauerToBessel}
\end{align}
We then find
\be
\int_0^M dp f(p) {\cal C}_{2,-p^2}^{\rm improved}[m^2,J]
 \sim \frac{2\Gamma\big(\tfrac{D-2}{2}\big)}{m^4}
 \int_0^M dp f(p)\ \frac{J_{\frac{D-4}{2}}(bp)}{(bp/2)^{\frac{D-4}{2}}}\quad\text{as}\quad m\rightarrow\infty\,,
 \label{flat fourier}
\ee
which is indeed proportional to the transverse-plane Fourier transform of $f(p)/p^{D-3}$, see \eqref{eq:impactparametertransform}.
This is a key finding:\\

\qquad\begin{minipage}{0.85\textwidth}{\it Bounds come from functions that have compact support in momentum space and are positive in impact parameter space.
}\end{minipage}\\

\noindent Fortunately, such functions are plentiful.
Note that in addition to imposing positivity in impact parameter space, we must impose positivity of \eqref{flat positivity} at finite $m\geq M$. This further restricts the space of possible functions $f(p)$, but we find numerically that suitable $f(p)$ still exist.

To obtain stronger bounds, we can supplement the ${\cal C}_{2,u}^\mathrm{improved}$ sum rule with additional null constraints 
\begin{align}
0 &= \left\< {\cal X}_{k,u}[m^2, J]\right\>, \qquad k=4,6,\dots
\end{align} 
where \cite{Caron-Huot:2020cmc}
\ba
\label{eq:nullconstraints}
{\cal X}_{k,u}[m^2,J] &= \frac{2 m^2+u }{u m^2 (m^2+u)}
\frac{m^2 \mathcal{P}_J\left(1+\frac{2 u}{m^2}\right)}{
 (u m^2 (m^2+u))^{k/2}
}-\\
& -
\Res_{u'=0}\frac{(2 m^2+u') (m^2-u') (m^2+2 u') }{ m^2 (u-u') u' (m^2-u) (m^2+u') (m^2+u+u')}
\frac{
m^2 \mathcal{P}_J\left(1+\frac{2 u'}{m^2}\right)
}{
( u' m^2(m^2+u'))^{k/2}
}.
\ea
${\cal X}_{k,u}$ can be derived by starting from $\cC_{k,u}$ and subtracting all the EFT contributions using the forward limit of $\cC_{k+2,u}$, $\cC_{k+4,u}$ etc. Including the ${\cal X}_{k,u}$ sum rules gives us the following linear program for bounding $g_2$ and $g_3$:
\ba
&\text{if}\quad\forall m\geq M,\ J=0,2,4\ldots\quad \int_0^M dp\, f(p) {\cal C}_{2,-p^2}^{\rm improved}[m^2,J ] + \\
&\qquad\qquad\qquad\qquad\qquad\qquad\qquad+\sum_{k=4,6,\dots} \int_0^M dp\,h_k(p) \mathcal{X}_{k,-p^2}[m^2,J] \geq 0\\
&\text{then}\quad\int_0^M dp\,f(p)\left[\frac{8\pi G}{p^2} +2g_2 +g_3p^2\right] \geq 0\,,
\label{eq:fulllinearprogram}
\ea
where the decision variables are the functions $f(p)$ and $h_k(p)$. We can choose an objective function and normalization condition to optimize different quantities.  For example, to obtain the best upper bound on $g_3$ as a function of $g_2$ and $8\pi G$, we must solve the otimization problem
\be
\mathrm{minimize}\int_0^M dp\,f(p)\left[\frac{8\pi G}{p^2} +2g_2\right]\qquad \textrm{such that }\int_0^M dp\,f(p)p^2 = -1,
\ee
where $f(p)$ and $h_k(p)$ satisfy the positivity constraints in (\ref{eq:fulllinearprogram}).

\subsection{Numerical implementation}

To solve the linear program (\ref{eq:fulllinearprogram}) numerically, we must express $f(p)$ and the $h_k(p)$ as finite sums of basis functions. Importantly, we must be able to find finite linear combinations of basis functions that are positive when Fourier-transformed to impact parameter space. It is well-known that polynomials restricted to an interval can have positive Fourier transforms.\footnote{An example is
\begin{align}
\int_0^1 (1-p) \cos(p b) dp&= \frac{1-\cos b}{b^2} \geq 0.
\end{align}
}
This motivates us to choose powers of $p$ as our basis functions, for example:
\begin{align}
f(p) &= \sum_{n} a_n p^n .
\end{align}
Here, $a_n$ are constants (decision variables) to be determined by solving the linear program, and $n$ are powers that we can choose. The powers need not be integers, but they must obey $n>1$ in order for the integral of $f(p)$ against the gravity pole $\frac{8\pi G}{p^2}$ to converge. For technical reasons that we explain in Appendix~\ref{sec:impactparamineqs}, we choose the basis functions listed in Table~\ref{tab:basisfunctionsforf}.\footnote{We have also investigated other choices of pure-power basis functions and found consistent results.} For the functions $h_k(p)$, we simply expand them in nonnegative integer powers: $p^0,p^1,\dots$. 
\begin{table}
\centering
\begin{tabular}{c|l}
$D$ & basis functions for $f(p)$ \\
\hline 
5 & $p^3-p^2,\,p^4-p^2,\,p^5-p^2,\,\dots$ \\
even $\geq 6$ & $p^{3/2},\, p^{5/2},\, p^{7/2},\,\dots$ \\
\phantom{j}odd $\geq 7$ & $p^2,\, p^3,\, p^4,\,\dots$ \\
\end{tabular}
\caption{Basis functions for $f(p)$ appearing in the linear program (\ref{eq:fulllinearprogram}). The technical reasons for these choices are explained in Appendix~\ref{sec:impactparamineqs}.
\label{tab:basisfunctionsforf}
}
\end{table}

We can now truncate the expansions of $f(p)$ and $h_k(p)$ in basis functions to obtain a linear program with a finite number of decision variables. We deal with the infinite number of constraints (one constraint for each $m>M$ and $J=0,2,4,\dots$) using a combination of discretization and polynomial approximation, described in detail in Appendix~\ref{app:flatspacenumerics}. Importantly, we must restrict $J$ to a finite range $J=0,2,\dots,J_\mathrm{max}$. To help the bounds converge without needing to take $J_\mathrm{max}$ very large, we explicitly include positivity in the scaling limit (\ref{flat fourier}) as an extra inequality.  We solve the resulting optimization problems numerically using {\tt SDPB} \cite{Simmons-Duffin:2015qma,Landry:2019qug}.

\subsection{Bounds on $g_3$ and $g_4$ without gravity}

\begin{figure}
\centering
\includegraphics[width=\textwidth]{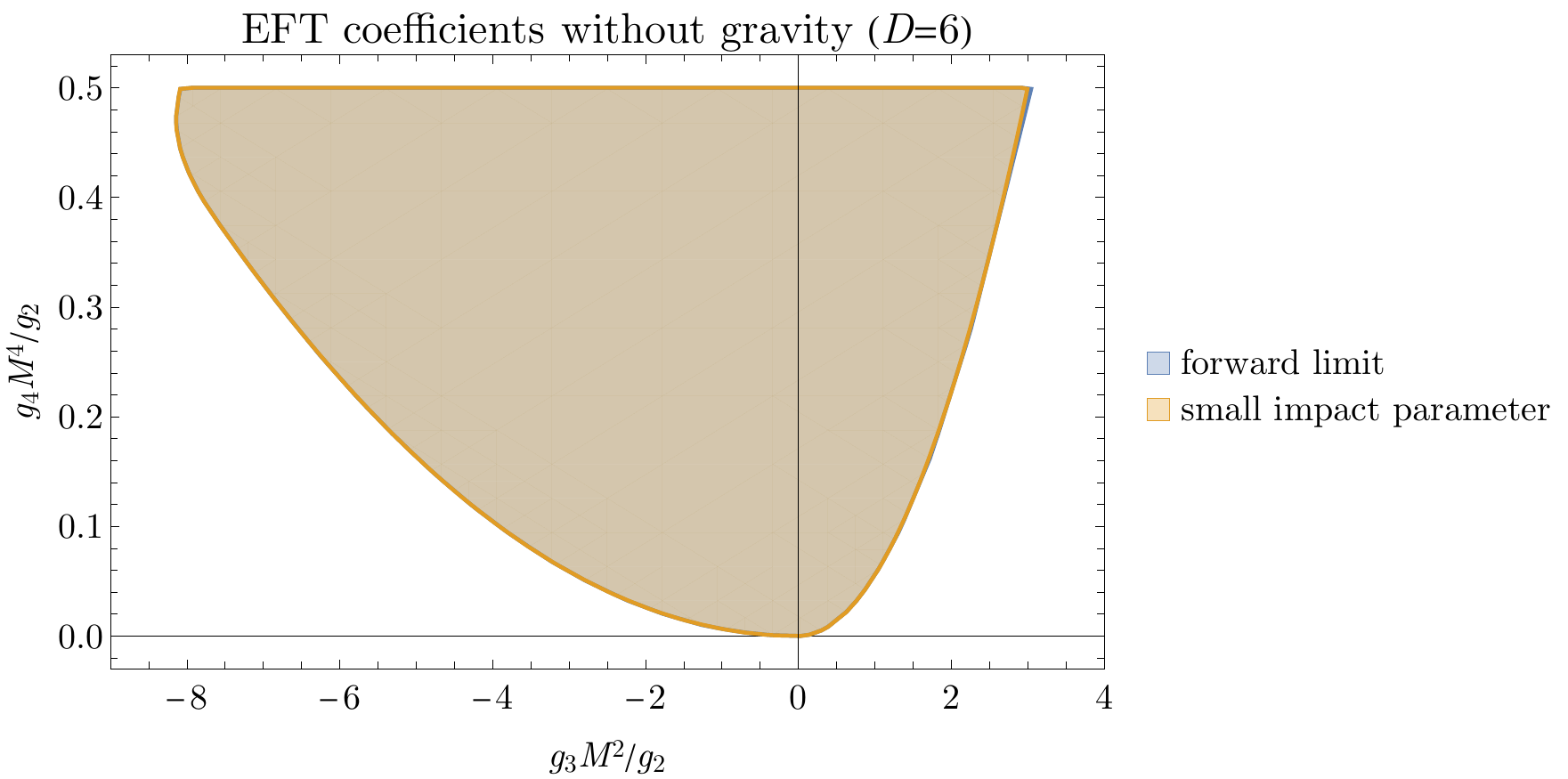}
\caption{Allowed region for $g_3$ and $g_4$ in a non-gravitational theory in $D=6$ dimensions, with heavy mass scale $M$. We show results using two different methods: the blue region uses derivatives around the forward limit as in \cite{Caron-Huot:2020cmc} (with a 33-dimensional space of functionals), while the yellow region uses small impact parameter wavepackets (built from the 17-dimensional space of functionals listed in Table~\ref{tab:parameters}, together with $\mathcal{C}_{4,u=0}$). The two regions are essentially identical and appear overlapping in the plot.  We give more details on our numerical computations in Appendix~\ref{app:flatspacenumerics}.
\label{fig:allowedregionsnogravity}
}
\end{figure}

In non-gravitational theories, dispersive bounds computed using the above methods reproduce the same results obtained by expanding around the forward limit. The physical reason is that heavy averages in dispersive sum rules are dominated by small impact parameters $b\sim 1/M$, as observed in \cite{Caron-Huot:2020cmc}. By using functionals that are localized on the scale $b\sim 1/M$, we access the same physics. As an example, in Figure~\ref{fig:allowedregionsnogravity}, we show a bound on $g_3,g_4$ in $D=6$, computed using our small impact parameter wavepackets. The results agree with those of \cite{Caron-Huot:2020cmc}, which used expansions around the forward limit.

\subsection{Bounds on $g_2$ and $g_3$ with gravity}

\begin{figure}
\centering
\includegraphics[width=0.9\textwidth]{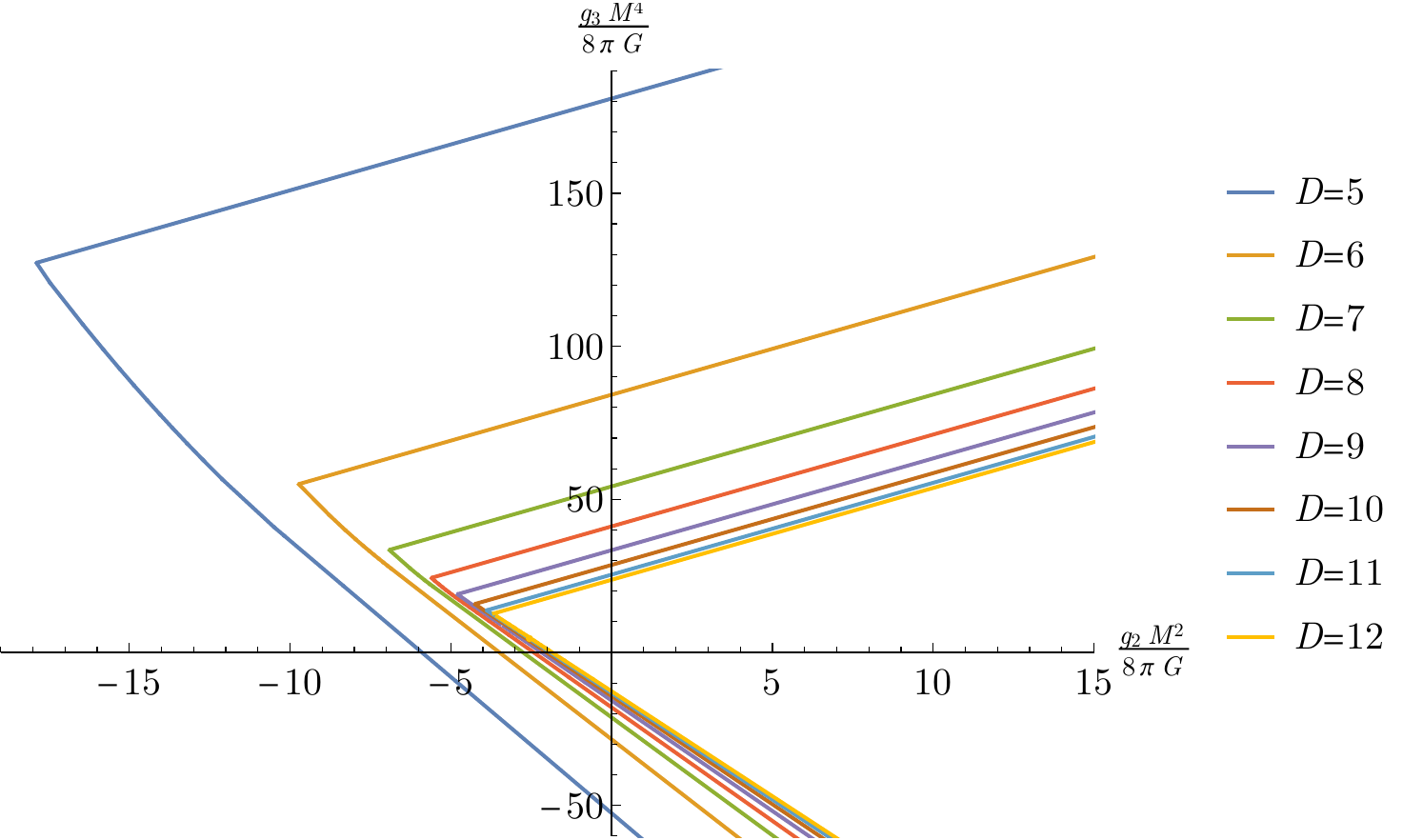}
\caption{Allowed regions for $g_2$ and $g_3$ in a theory of a scalar coupled to gravity in flat space in dimensions $D=5,\dots,12$, with heavy mass scale $M$. For each curve, the region to the right is allowed and the region to the left is disallowed. Each bound was computed using a 17-dimensional space of functionals, listed in Table~\ref{tab:parameters}. We give more details on the numerical computation in Appendix~\ref{app:flatspacenumerics}. The inequalities plotted here are listed in Table~\ref{eq:inequalitiesforfigure1}.
\label{fig:allowedregionsdifferentD}
}
\end{figure}

The main advantage of our approach is that we also obtain valid bounds in gravitational theories. In Figure~\ref{fig:allowedregionsdifferentD}, we show the allowed regions for $g_2$ and $g_3$ in UV-completable tree-level EFTs containing gravity in spacetime dimensions $D=5,\dots,12$. (We discuss the case $D=4$ in Section~\ref{sec:dim4}). The bounds are computed numerically by solving (\ref{eq:fulllinearprogram}) using {\tt SDPB}. Note that they automatically have the expected EFT scaling in $M$. In short, dimensional analysis scaling is a theorem, {\it even in the presence of gravity!}

The bounds in Figure \ref{fig:allowedregionsdifferentD} are computed using a 17-dimensional space of functionals built from ${\cal C}_2^\mathrm{improved}$ and the null constraints $\mathcal{X}_4$ and $\mathcal{X}_6$ (listed explicitly in in Table~\ref{tab:parameters}). Although the bounds depend on the cutoffs and approximations described in Appendix~\ref{app:flatspacenumerics}, we have chosen those cutoffs so that the results have converged within a fraction $10^{-4}$. The bounds can be improved by choosing a larger space of functionals. We expect that the bounds shown in Figure~\ref{fig:allowedregionsdifferentD} are within a few percent of optimal.\footnote{Here, we mean ``optimal'' for the specific infinite dimensional linear program (\ref{eq:fulllinearprogram}). One could potentially obtain stronger bounds by making new assumptions about the theory in question, or by extending the range of $u$ as discussed in Appendix~\ref{app:extendedu}.}

In Figure~\ref{fig:wavepacket}, we show the impact parameter wavefunction $\widehat f(b)$ for the extremal functional that minimizes $g_2$ in $D=6$. Clearly, the numerical optimization procedure
constructs sum rules dominated by $b\sim 1/M$.

\begin{figure}
\centering
\includegraphics[width=0.7\textwidth]{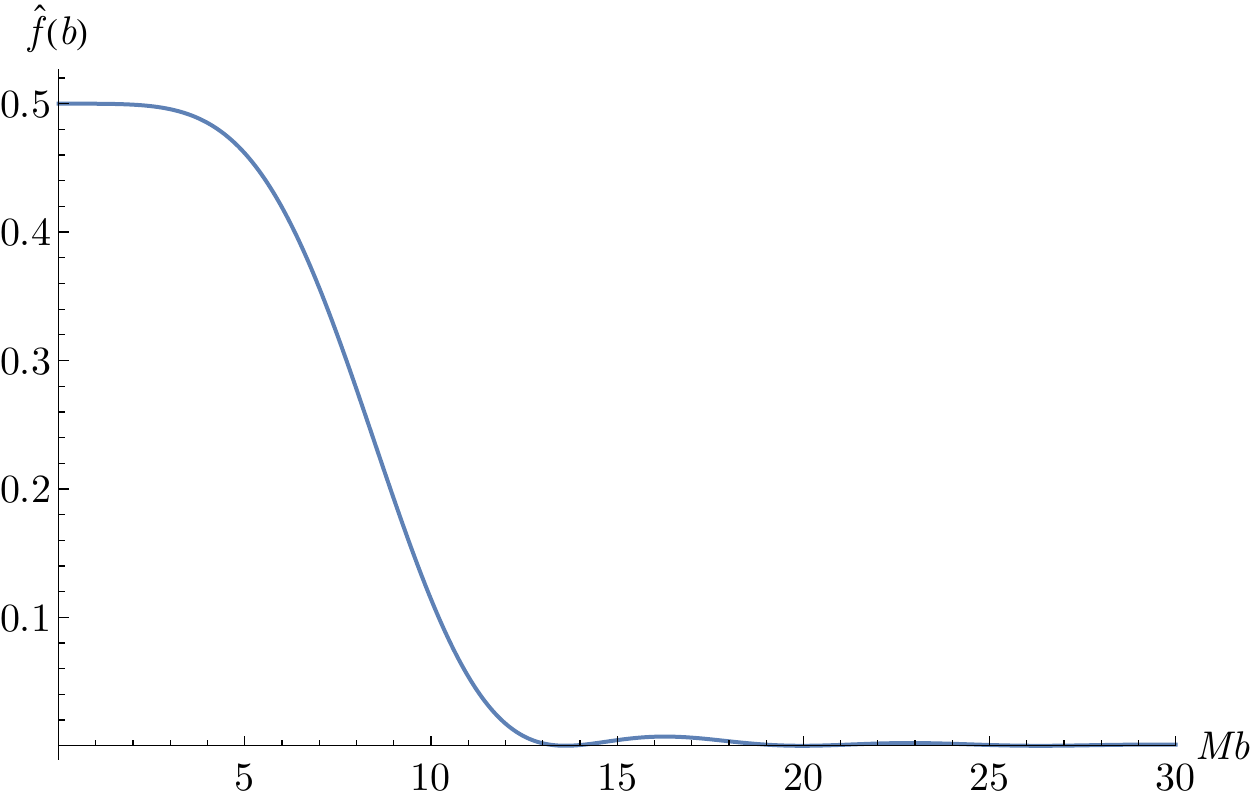}
\caption{The impact parameter wavefunction $\widehat f(b)$ defined by (\ref{eq:impactparametertransform}) for the extremal functional that minimizes $g_2$ in $D=6$. As discussed in the text, it is localized near $b\sim 1/M$. The wavefunction is normalized by $\widehat f(0)=1/2$ so that the contribution of $g_2$ in (\ref{eq:fulllinearprogram}) is precisely $g_2$. The fact that it has zero slope at $b=0$ guarantees that the contribution of $g_3$ vanishes.
\label{fig:wavepacket}
}
\end{figure}

Because our sum rules are linear and homogeneous in the EFT couplings $8\pi G, g_2, g_3$, we can always add an admissible amplitude without gravity to an admissible amplitude with gravity to obtain a new admissible amplitude with gravity. The allowed region in $(g_2,g_3)$-space without gravity is a cone $C$ \cite{Caron-Huot:2020cmc}. The allowed region with gravity must be a union of translations of $C$. Indeed, this is the case: the allowed region is similar to the non-gravitational one, but shifted so that $g_2$ has a negative minimum value (achieved at a particular value of $g_3$). Note that the bounds are stronger in larger $D$. This is due to the fact that the dimensional reduction of a unitary theory is unitary (more technically the fact that higher-dimensional Gegenbauer polynomials can be written as positive linear combinations of lower-dimensional Gegenbauer polynomials).
Physically, it makes sense that the ratio  $g_2 M^2 / G$ should not admit an upper bound: $g_2$ and $G$ are  a priori independent couplings, measuring respectively the strength of the scalar self-interaction and the strength of gravity. We are assuming that  the EFT is weakly coupled, which means that {\it both} $g_2$ and $G$ are taken to be small in units of $M$, but their ratio is  a priori undetermined without further physical input. On the other hand, for fixed $g_2 M^2 / G$, we expect (and will confirm) that all other dimensionless ratios $g_k M^{2k-2}/G$  obey double-sided bounds.

\begin{figure}
\centering
\includegraphics[width=0.9\textwidth]{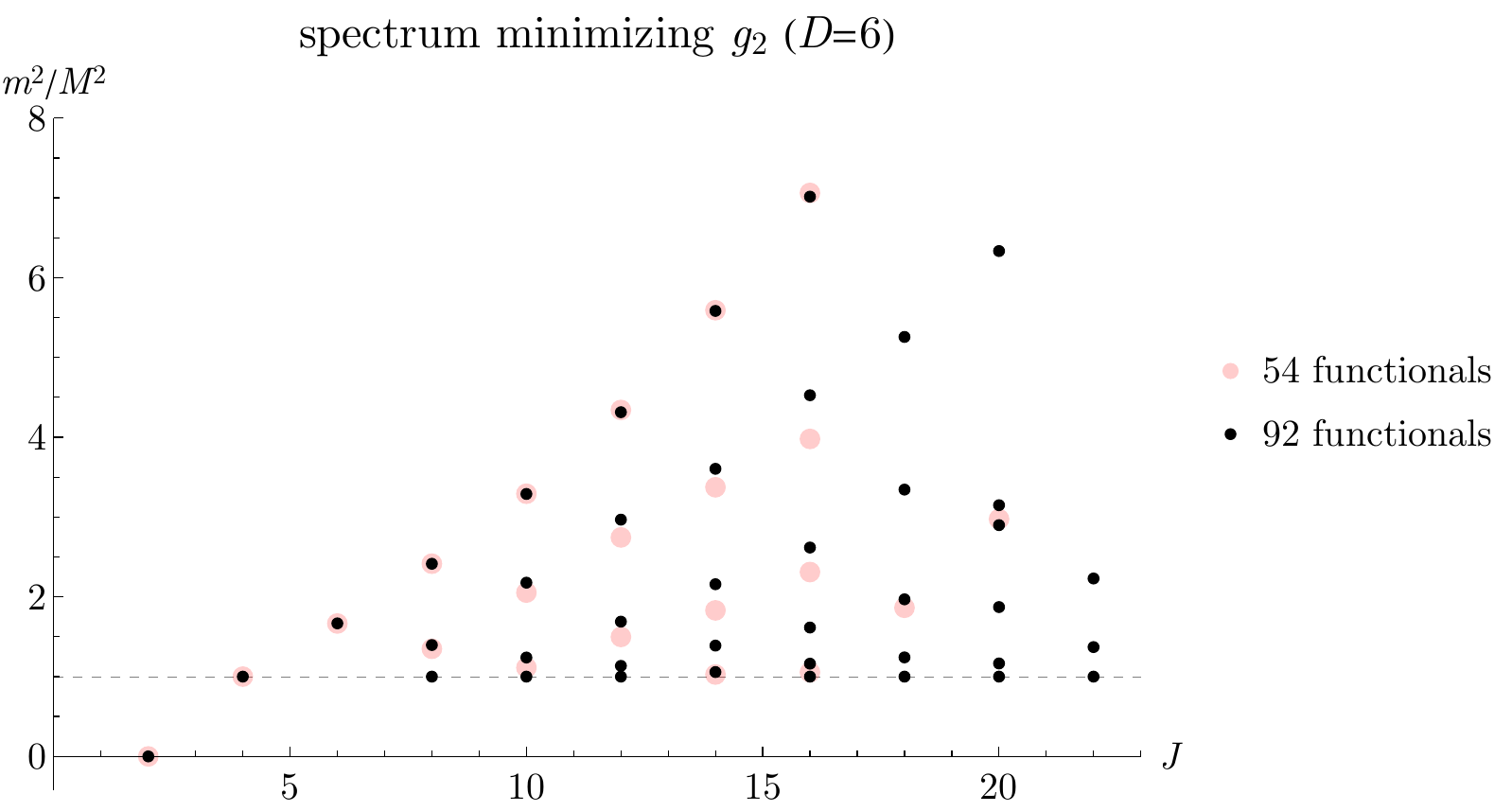}
\caption{Extremal spectrum for the problem of minimizing $g_2$ in spacetime dimension $D=6$. The black points are zeros of the extremal functional, computed using a 92-dimensional space of functionals described in Table~\ref{tab:parameters}. (We only show zeros that are stable under changing the space of functionals and other parameters.) For comparison, we show the spectrum computed using a 54-dimensional space of functionals in red. The upper trajectory has mostly converged, but the lower trajectories are still changing as we increase the space of functionals. The extremal value of $g_2$ corresponding to this spectrum is $g_2M^2/(8\pi G) = -9.57$. In string theory, the spectrum would occupy the upper triangle, but here it is in the lower triangle.
\label{fig:extremalspectrum}
}
\end{figure}

The slope of the upper bound on $g_3$ as a function of $g_2$ is exactly 3. This comes from the fact that the scalar contribution to the ${\cal C}_2^\mathrm{improved}$ sum rule is
\begin{align}
{\cal C}_{2,-p^2}^\mathrm{improved}[m^2,J=0] &= \frac{2}{m^4} + \frac{3 p^2}{m^6},
\end{align}
which has the same form as the low-energy contribution $2g_2+g_3 p^2$ in (\ref{eq:fulllinearprogram}). By adding scalars  to the heavy spectrum, we can shift $(g_2,g_3)$ by an arbitrary positive multiple of $(1,3/m^2)$. The minimum value of $g_2$ is achieved by a spectrum with no heavy scalars. We can compute this spectrum using the extremal functional method \cite{Poland:2010wg,ElShowk:2012hu}: we find functions $f(p)$ and $h_k(p)$ that give the optimal lower-bound on $g_2$. We then tabulate the values of $m^2$ and $J$ where the inequalities in (\ref{eq:fulllinearprogram}) are saturated --- i.e.\ the zeros of the extremal functional. We show the resulting spectrum for the case $D=6$ in Figure~\ref{fig:extremalspectrum}. The extremal spectrum is remarkable (and very different from string theory) in that there is only a single state at spins $J=2,4,6$, and a small but increasing number of states at larger $J$. Furthermore, the minimal value of $m^2$ appears to be nearly flat as a function of $J$. It is interesting to ask whether there could be a physical theory of gravity that realizes this spectrum.

\subsection{Bounds on higher contact coefficients with gravity}

\begin{figure}
\centering
\includegraphics[width=\textwidth]{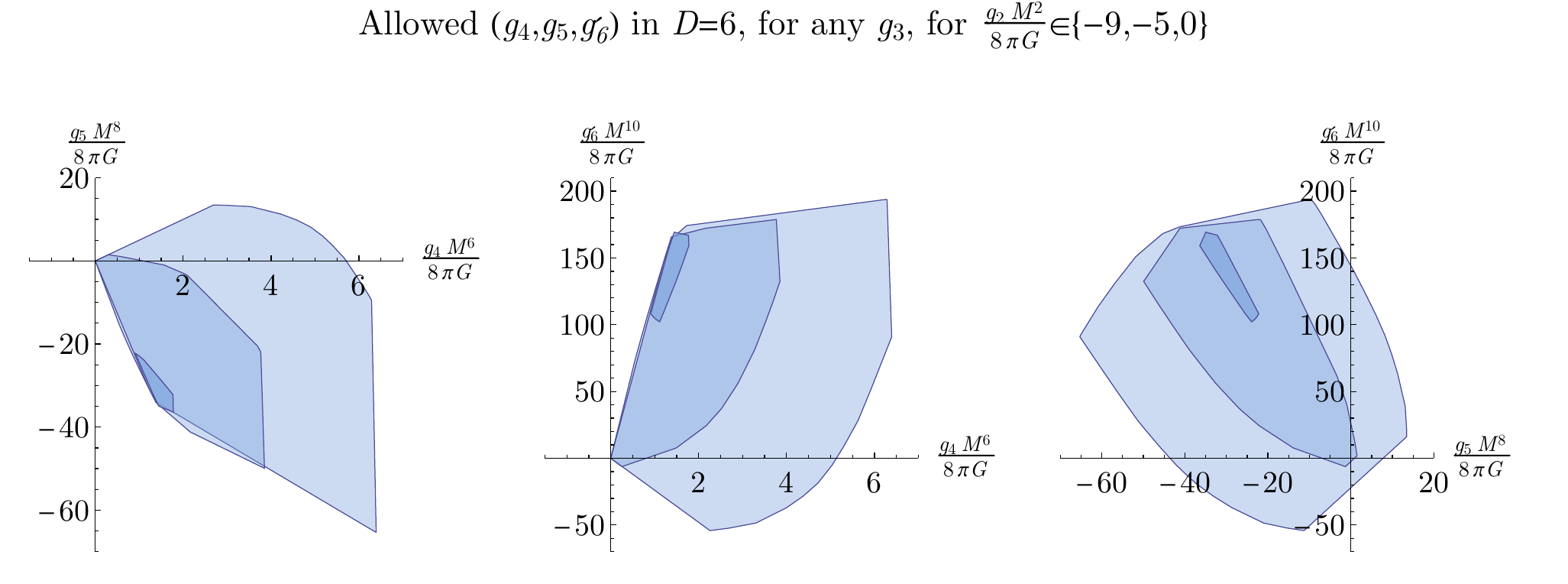}
\caption{
Bounds on the higher-dimension contact coefficients $g_4,g_5,g^\prime_6$ in $D=6$. We compute bounds that are independent of $g_3$ by restricting to the subspace of functionals whose EFT contribution is independent of $g_3$. We compute the 3-dimensional allowed region in $(g_4,g_5,g^\prime_6)$-space for a few example values of $g_2$, and then project onto pairs of axes for display. The values of $g_2$ shown here are $g_2=0$ (lightest blue), $\frac{g_2 M^2}{8\pi G}=-5$ (medium blue), $\frac{g_2 M^2}{8\pi G}=-9$ (darkest blue). These bounds were computed using the same parameters and functionals listed in column 1 of Table~\ref{tab:parameters}, with the additional functionals $\cC_{4,-p^2}^\mathrm{improved}\x\{p^0,p^1,\dots,p^5\}$. When $g_2$ approaches the ``tip of the cone'' at its minimum value $g_2M^2/(8\pi G) \approx -9.6$, the allowed region shrinks towards $\p{\frac{g_4 M^6}{8\pi G}, \frac{g_5 M^8}{8\pi G},\frac{g^\prime_6 M^{10}}{8\pi G}}\approx (1.4,-34,165)$.
\label{fig:highercontacts}
}
\end{figure}

The same method straightforwardly extends to higher EFT coefficients. Using the same strategy as in (\ref{eq:definitionofb2improved}), we can define a ${\cal C}_{4, u}^\mathrm{improved}$ sum rule that isolates the coefficients $g_4,g_5,g^\prime_6$:
\ba
&4g_4 - 2u g_5 + g^\prime_6 u^2 = \left\<{\cal C}_{4, u}^\mathrm{improved}[m^2,J]\right\> \\
&{\cal C}_{4, u}^\mathrm{improved}[m^2,J] = 
\frac{(2 m^2+u) \mathcal{P}_J(1+\frac{2 u}{m^2})}{m^4 (m^2+u)^3}
-\frac{u^2 (6 m^6-4 m^2 u^2+4 m^4 u-3 u^3) \mathcal{P}_J(1) }{m^{12} (m^2+u)^3} \\
\quad
&\qquad\qquad\qquad\qquad-\frac{2 u^3 (-3 m^2 u+4 m^4-5 u^2) \mathcal{P}_J'(1)}{m^{12} (m^2-u) (m^2+u)^2}
+\frac{4 u^4 \mathcal{P}_J''(1)}{m^{12} (u-m^2) (m^2+u)}.
\ea
The coefficient $g_6$, which multiplies $(s^2+t^2+u^2)^3$ in the Lagrangian, was eliminated
since it can be measured by the higher-subtracted sum rule ${\cal C}_{6,u}$:
we will therefore not discuss it here.
We then generalize (\ref{eq:fulllinearprogram}) to include linear combinations of the ${\cal C}_{2,-p^2}^\mathrm{improved}$, ${\cal C}_{4,-p^2}^\mathrm{improved}$ and ${\cal X}_{k,-p^2}$ sum rules, each integrated against its own function of $p$.\footnote{Alternatively, since the forward limits $u\to 0$ of spin-4 sum rules converge, we could simply use derivatives of the un-improved ${\cal C}_4$, supplemented with null constraints, as opposed to using ${\cal C}_{4, u}^\mathrm{improved}$.}  By finding linear combinations that are positive for all $m\geq M$ and $J=0,2,4,\dots$, we obtain inequalities on EFT data with the correct scaling in $8\pi G$ and $M$. As an example, in Figure~\ref{fig:highercontacts} we show bounds on $g_4,g_5,g^\prime_6$ in spacetime dimension $D=6$, for some example values of $g_2$.

\subsection{Some solutions to the constraints from string theory}

It is interesting to compare our bounds to explicit solutions of crossing symmetry with gravity. The exchange of gravitons in the three channels alone violates the Regge bound \eqref{eq:ReggeBound} and needs to be UV completed. A simple UV completion is provided by the amplitude of four real dilatons in type II string theory \cite{Schwarz:1982jn}
\be
\MM_{\text{string}}(s,t)= -8\pi G \frac{(s t)^2+(t u)^2+(s u)^2}{m^6}\frac{\Gamma\left(-\tfrac{s}{m^2}\right)\Gamma\left(-\tfrac{t}{m^2}\right)\Gamma\left(-\tfrac{u}{m^2}\right)}{\Gamma\left(\tfrac{s}{m^2}+1\right)\Gamma\left(\tfrac{t}{m^2}+1\right)\Gamma\left(\tfrac{u}{m^2}+1\right)}\,.
\label{eq:MString}
\ee
Here $m\geq M$ is the mass of the lightest massive state exchanged. It is straightforward to check that along any ray of constant phase in the $s$ plane away from the real axis, at fixed $u$, we have
\be
|\MM_{\text{string}}(s,u)| = O(|s|^{2+2u/m^2})\,,
\ee
in agreement with \eqref{eq:ReggeBound}. Furthermore, the amplitude is unitary for $2\leq D \leq 24$.\footnote{Unitarity for $2\leq D\leq 10$ follows from unitarity of type II superstring theory. The residues of \eqref{eq:MString} at the massive poles experimentally admit a positive expansion into Gegenbauer polynomials in the extended range $2\leq D\leq 24.148\ldots$. We are not aware of a proof of unitarity of \eqref{eq:MString} in the extended range, see \cite{NimaTalkStrings2016}.} The low-energy expansion of \eqref{eq:MString} starts as follows
\be
\MM_{\text{string}}(s,t) = 8\pi G\left(\frac{tu}{s}+\frac{us}{t}+\frac{st}{u}\right) + \frac{4\pi G\zeta(3)}{m^6}(s^2+t^2+u^2)^2+\ldots,
\ee
i.e. this amplitude has
\be
g_2 = 0\,,\qquad g_3 = 0\,,\qquad g_4=  \frac{4\pi G\zeta(3)}{m^6}\,,
\ee
meaning that the point
\be
\frac{g_2M^2}{8\pi G} = 0\qquad \frac{g_3M^4}{8\pi G} = 0\qquad \frac{g_4M^6}{8\pi G} = x \frac{\zeta(3)}{2}
\ee
is consistent with all the constraints for any $0 < x\leq 1$ as long as $2\leq D\leq 24$.

This point lies at the origin in Figure \ref{fig:allowedregionsdifferentD}
and does not come close to saturating the bounds on $g_2$ and $g_3$.
One can construct a consistent solution of the constraints lying closer to the bound by subtracting all the scalar (spin-0)
exchanges from $\MM_{\text{string}}$
but this only has the effect of translating the solution
to the lower-left of the origin in Figure \ref{fig:allowedregionsdifferentD}. For example, in $D=6$ we find:
\be
\frac{g_2 m^2}{8\pi G} \approx -0.75\,,\qquad \frac{g_3 m^4}{8\pi G} \approx -2.07\,,
\ee
bringing us somewhat closer to the numerical bound, but still far from saturation. We can also consider scattering of real dilatons in the heterotic string \cite{Kawai:1985xq}, where we find
\be
\frac{g_2 m^2}{8\pi G} = \frac{3}{16}\,,\qquad \frac{g_3 m^4}{8\pi G} = \frac{3}{4}\,,
\ee
also safely within our bounds.

String theory is of course by no means the unique way to ``unitarize'' tree-level two-to-two
graviton exchange, especially at large impact parameters.
Perhaps the simplest idea (pre-dating string theory) is to exponentiate the eikonal phase.
To see this let's transform the amplitude to impact parameter space as
\be
 \mathcal{M}(s,b)\equiv \frac{1}{2s} \int \frac{d^{D-2}p}{(2\pi)^{D-4}} e^{ip{\cdot}b} \mathcal{M}(s,u=-p^2)
\ee
which allows to write the eikonal amplitude as (see \cite{Amati:1987wq} in the context of string theory)
\be \mathcal{M}_{\rm eik}(s,b) = -i\left(e^{is\hat{\chi}(b)}-1\right)\,,\qquad \hat{\chi}(b) = \frac{G\Gamma(\tfrac{D-4}{2})}{(\pi b^2)^{\frac{D-4}{2}}}\,.
\ee
Here we are thinking of $b$ large compared to the string scale and other scales.
Expanding the exponential to first order reproduces tree-level graviton exchange.
It is not hard to see that this model indeed saturates the
${\cal C}_{2, u}^{\rm improved}$ sum rule from eq.~\eqref{B2 improved flat},
with the lower endpoint removed: in Fourier space that sum rule reads
\be
 \hat{\chi}(b) = \frac{2}{\pi} \int_{0}^\infty \frac{ds}{s^2} {\rm Im} \mathcal{M}(s,b)
 \to_{\rm eik} \frac{2}{\pi} \int_0^\infty \frac{ds}{s^2} 2\sin^2(\frac{s\hat{\chi}(b)}{2})\,.
\label{Im M eik}
\ee
The eikonal model does not quite satisfy the assumptions in the paper and so we do not
put the corresponding data in the same plot (one would have to choose a scheme for
subtracting the low-energy loop contribution to the spectral density from $s<M^2$), however this contribution is negligible at sufficiently large impact parameters,
on which we will focus here.
One can interpret the above sum rules as constraining high-energy data at fixed ``impact parameter'' $b=2J/m$. This may be seen using eqs.~\eqref{eq:UVaverage} and \eqref{eq:GegenbauerToBessel} and integrating using orthogonality of Bessel functions:
\be
  \hat{\chi}(b) \approx \avg{ \frac{1}{bm^4} \delta\left(b-\frac{2J}{m}\right)}
 \frac{\Gamma(\tfrac{D-2}{2})}{(b/2)^{D-4}}\,. \label{large b sum rule 2}
\ee
This should be understood in the sense of distributions, with sufficient smearing
in $b$ to satisfy the finite-$p$ support condition, and it shows that low-energy gravity predicts high-energy averages at fixed $J/m$.
These constraints are of course satisfied in tree-level string theory, but rather differently
than in the eikonal approximation.
Due to the famous logarithmic spreading with energy,
\be
 {\rm Im}\ \mathcal{M}_{\rm string}(s,b)  \propto e^{-\frac{b^2M^2}{8\log s}} \label{Im M string}
\ee
at large impact parameters $bM\gg 1$ string theory's spectral density effectively
vanishes below \emph{exponentially} large $s$.

The spectral densities
(\ref{Im M eik}) and (\ref{Im M string}) could hardly differ more from each other.
Yet they are both physically reasonable and are both realized in different regimes of string theory.
In our view, this indicates that high-energy models should be used with great care
when deriving EFT bounds, or perhaps avoided (as we do in this paper). 

\subsection{Comments on $D=4$ and infrared divergences}
\label{sec:dim4}

Trying to play the above game for gravitational theories in $D=4$, we run into a problem. In order for the integral against gravity to converge, we need $\lim_{p\to 0} f(p)/p$ to vanish. However, when $D=4$, this limit is proportional to $\int d^2 \vec b \hat f(\vec b)$, which must be strictly positive.

In other words, if we restrict to functionals that are positive at all impact parameters, the action on gravity is logarithmically divergent.
This reflects the infrared divergence of the gravitational potential in the Regge limit, given by the
two-dimensional Fourier transform $\int \frac{d^2p}{p^2} e^{ip{\cdot}b}$.

A simple solution is to require positivity only at distances less than an IR cutoff $b<b_{\rm max}$.
In practice, this can be achieved simply by introducing a small cutoff in momentum space,
$|p|>p_{\rm min}$ where $p_{\rm min}\ll M$.
We then consider the linear programming problem in \eqref{eq:fulllinearprogram} but replacing the action on gravity
by $f_{2}(p=0)$, which gives the coefficient of the logarithmic divergence $\sim\log\frac{1}{p_{\rm min}}$.
In this ``leading-log'' approximation the shape of the extremal functional is independent of the cutoff.  The cutoff $p_{\rm min}$ can then be quantitatively related to $b_{\rm max}$,
the impact parameter at which the action becomes negative, by plotting the Fourier transform of the functional. In fact, the relation can be found analytically by series-expanding at small $p_{\rm min}$, from which we find $p_{\rm min}^2 =c/(Mb_{\rm max}^3)$
where the numerical constant $c\approx 1$ is found from the extremal functional. 
In this way we obtain the lower bound:\footnote{An earlier arXiv version of this paper
used an incorrect relation between $p_{\rm min}$ with $b_{\rm max}$, which resulted in a numerically incorrect bound.}
\be
 g_2 \geq -\frac{8\pi G}{M^2} \times  25\log (0.3 Mb_{\rm max})\, \qquad (D=4)\,.
\ee

What physical value should we choose for the cutoff?
In the context of AdS/CFT, there will be a clear choice: the AdS scale $b_{\rm max}\sim R_{\rm AdS}$.
But this could be an over-estimate since the
distance $b_{\rm max}$ need simply be a scale outside which we consider
the amplitude to be computable, for example using the eikonal approximation.
Negativity of a functional is not necessarily a problem if it occurs in a region under analytic control.
In Minkowski space, it should be possible to make this analysis fully rigorous by
considering coherent states of the scalar and its radiation;
this would require knowing the properties of such dressed states under crossing symmetry and discontinuities. We leave this to future work.

\subsection{Maximal supergravity: bounding graviton scattering}

Can two gravitons produce heavy states with an arbitrary cross-section, or must all processes
involving gravitons be suppressed by $G$?
Here we give partial support for the latter idea,
in the special case of maximal supersymmetry.

The technical simplification is that the graviton lies in the same multiplet as a scalar.
We can factor out the helicity dependence and effectively we have a massless real scalar with low-energy amplitude
\be
\mathcal{M}_{\text{susy}}(s,u)  = \frac{8\pi G}{s t u} + g_0 + g_2 (s^2+t^2+u^2) + \ldots\,.
\ee
Maximal supersymmetry effectively improves the $s\to\infty$ behavior by four powers, so that we have the high energy bound
\be
s^2\mathcal{M}_{\text{susy}}(s,u)\rightarrow 0\quad\text{as}\quad s\rightarrow\infty\,.
\ee
The $(-2)$-subtracted dispersion relation now converges for $u<0$.
The corresponding ${\cal C}_{-2}$ and ${\cal C}_0$ sum rules read:
\begin{align}
{\cal C}_{-2} &:\ \frac{8\pi G}{-u} =
 \avg{m^2(2m^2+u)\legP_J(1+\frac{2u}{m^2})}\,, \label{B-2 sugra} \\
{\cal C}_0& :\ g_0 + 2 g_2u^2 + \ldots =
\avg{\frac{(2m^2+u)\legP_J(1+\frac{2u}{m^2})}{m^2+u}}\label{B0 sugra}\,.
\end{align}
Note how gravity enters its own sum rule, apparently decoupled from the rest. This is a simplifying feature of supersymmetry.
Since the second line involves higher powers of $1/m^2$ than the first, we can bound
$g_0$ in terms of gravity and the heavy mass $M$. Again the trick is to measure $g_0$ using small impact parameters.

Proceeding as in \eqref{B2 improved flat}, we may eliminate the higher contacts $2g_2u^2+\ldots$ from the left-hand side of \eqref{B0 sugra} using higher subtracted sum rules.
A shortcut is to use $s\leftrightarrow u$ symmetry of the unsubtracted dispersion relation:
\be
 \avg{ \frac{m^2(2m^2+u)\legP_J(1+\tfrac{2u}{m^2})}{(m^2-s)(m^2+s+u)}}
  =\avg{\frac{m^2(2m^2+s)\legP_J(1+\tfrac{2s}{m^2})}{(m^2-u)(m^2+s+u)}} \qquad(s,u<0)\,.
\ee
Note that this equality involves only heavy data: the massless poles contribute $\frac{1}{stu}$
to both sides and cancel each other.  Setting $s=0$ we obtain a family of null constraints
\be
0=\avg{\frac{(2m^2+u)\legP_J(1+\frac{2u}{m^2})}{m^2+u} - \frac{2m^4}{m^4-u^2}} \qquad (u<0)\,,
\ee
where the first term is the average that previously appeared in (\ref{B0 sugra}).
Adding the identity $g_0=\langle2\rangle$ then gives the desired analog of \eqref{B2 improved flat}:
\be\begin{aligned}
 g_0 &= \avg{ \frac{(2m^2+u)\legP_J(1+\frac{2u}{m^2})}{m^2+u} - \frac{2u^2}{m^4-u^2} }_{\rm heavy}\,.
\\ &\equiv \avg{{\cal C}_{0,u}^{\rm improved}[m^2,J]}\,.
\label{g0 sum rule for any u}
\end{aligned}\ee
Compared with \eqref{B0 sugra}, the left-hand-side is now under complete control.
We have an infinite family of ways to measure $g_0$, labelled by $-M^2<u\leq 0$.

Using the linear programming strategy in \eqref{eq:fulllinearprogram}, now integrating the gravity measurement
in \eqref{B-2 sugra} against powers $p^2,p^3,\ldots p^{10}$, the $g_0$ measurement in \eqref{g0 sum rule for any u}
against four powers of $p$, and ${\cal X}_2$ against two powers of $p$ (for a total of 14 functionals),
we proved the following bound:
\be
 0\leq g_0 \leq 3.000 \frac{8\pi G}{M^6}\ \qquad\mbox{($D=10$, maximal supergravity)\,.}
 \label{eq:SUGRAresult}
\ee
We find it remarkable that such a bound exists at all.  It shows that in a theory with 32 real supercharges
all interactions must shut down as $G\to 0$.
It is an interesting question whether a similar statement holds for the coupling of gravitons
to heavy states in non-supersymmetric theories.

The bound is compatible with type II string theory, where
\be
\frac{g_0M^6}{8\pi G} =  2\zeta(3)\approx 2.40<3.000\,.
\ee
It would be interesting to find a model which saturates the bound.
Note that the bound is sharp only for theories weakly coupled below the scale $M$ (i.e.\ $M\ll M_{\rm pl}$),
since we neglected EFT loops.

Recent work \cite{Guerrieri:2021ivu} also considered bounds on $g_0$ with maximal supersymmetry in ten dimensions. Their set up differs from ours in that they consider bounds on $g_0/8\pi G$ not in the units of the UV cutoff $M$ but in Planck units $M_{\text{pl}}$, without assuming weak coupling $M\ll M_{\text{pl}}$. Therefore, in their case there is no upper bound on $g_0 M^6_{\text{pl}}/8\pi G$ since this ratio gets arbitrarily large in weakly coupled string theory. Intriguingly, they provide evidence that besides the rigorous bound $g_0\geq 0$, there should be a stronger lower bound of the form $g_0 M^6_{\text{pl}}/8\pi G\geq c>0$, possibly saturated by strongly coupled string theory. In the units of the UV cutoff, this lower bound approaches zero at weak coupling and thus is compatible with \eqref{eq:SUGRAresult}.


\section{Conclusions}\label{sec:conclusions}

In this paper, we considered higher derivative corrections in UV consistent gravitational theories in flat space. We explained how to derive bounds on these corrections using dispersion relations for the S-matrix. We focussed on the weakly coupled regime, meaning that the gravitational and any other interaction is very small. In this regime, the ratios $g_n/8\pi G$ of higher derivative couplings to the gravitational coupling are fixed numbers. On physical grounds, it is expected that in consistent theories, these numbers should be suppressed by inverse powers of the UV cutoff $M$. Here we define $M$ to be the mass of the first state which does not appear in the low-energy effective field theory. It has been known for some time~\cite{Camanho:2014apa} that theories where higher derivative corrections are large in the units of $M$ violate causality. Nevertheless, the long-standing challenge has been to turn such parametric bounds into precise bounds on the order one coefficients. In this paper, we derived such bounds.

We solved the problem in the context of theories containing a light (massless) scalar coupled to gravity. In such theories, we can consider higher derivative contact self-interaction of the scalar particle. The leading interaction, which we denoted $g_2$, has four derivatives, followed by $g_3$ with six derivatives, $g_4$ with eight etc. In \emph{non-gravitational} theories, causality and unitarity have long been known to imply that $g_2$ must be positive \cite{Adams:2006sv}. More recently, $g_3/g_2$, $g_4/g_2$ etc. have been argued to satisfy two-sided bound in the units of $M$ in the absence of gravity \cite{Tolley:2020gtv, Caron-Huot:2020cmc}, starting from dispersion relations expanded around the forward limit. However, the incorporation of gravity poses an obstruction to such program since the exchange of massless gravitons gives rise to a pole at vanishing momentum transfer.

In this paper, we overcame this difficulty by localizing the dispersion relations at small impact parameters rather than at small momentum transfer. This automatically leads to the correct EFT scaling and gives rise to a robust bootstrap program for bounding the order one coefficients. We showed that $g_2$ is bounded from below and that $g_3$, $g_4$ and higher couplings are confined to compact regions at a fixed $g_2$, see Figures \ref{fig:allowedregionsdifferentD} and \ref{fig:highercontacts}. The theory minimizing the $g_2$ coupling exhibits a peculiar spectrum shown in Figure \ref{fig:extremalspectrum}. It would be interesting to understand if it can come from a consistent theory of gravity.

We also considered theories with maximal supersymmetry. In this case, we were able to give both upper and lower bound on the leading correction, corresponding to $R^4$. This shows that all interactions must shut down in a theory of gravity with maximal supersymmetry if we take $G\rightarrow 0$.

Our results opens up several obvious avenues for future research. The type of question addressed here in flat space has a natural analogue in AdS, and will be the subject of an upcoming work \cite{CHMRSD3}. It would also be extremely interesting to derive similar bounds in the presence of a positive cosmological constant. To do that, one would first need to clarify the consequences of causality and unitarity in de Sitter space.

We have focussed on the simplest example of identical massless scalars. It will be natural to consider in our framework the scattering of more general external states -- most fundamentally, graviton scattering. We have already mentioned the need for a refinement of our method in $D=4$ to handle the IR divergence in the impact parameter representation. The incorporation of EFT loops is another natural direction. Our results can be regarded as a step in the classification program of weakly coupled theories of gravity, in the spirit of~\cite{Camanho:2014apa}.  The tools are now mature to pursue this program systematically.

\pagebreak
\section*{Acknowledgments}
The authors would like to thank Walter Landry, Yue-Zhou Li and Julio Parra Martinez for useful discussions. 
The work of S.C.-H. is supported by the
National Science and Engineering Council of Canada, the Canada Research Chair program,
the Fonds de Recherche du Qu\'ebec -- Nature et Technologies, the Simons Collaboration on
the Nonperturbative Bootstrap, and the Sloan Foundation. D.M. gratefully acknowledges funding provided by Edward and Kiyomi Baird as well as the grant DE-SC0009988 from the U.S. Department of Energy. The work of L.R. is supported in part by NSF grant \# PHY-1915093 and by the Simons Foundation (Simons Collaboration on the Nonperturbative Bootstrap and Simons Investigator Award). D.S.-D. is supported by Simons Foundation grant 488657 (Simons Collaboration on the Nonperturbative Bootstrap) and a DOE Early Career Award under grant no. DE-SC0019085. Some of the computations in this work were performed on the Caltech High-Performance Cluster, partially supported by a grant from the Gordon and Betty Moore Foundation.

\appendix


\section{Details on numerics}
\label{app:flatspacenumerics}

In this appendix, we give details on our numerical implementation of the linear program (\ref{eq:fulllinearprogram}), which we reproduce here for convenience.
\begin{align}
\label{eq:thepositivityconditions}
\mbox{if: }& \int_0^1 dp\, f(p) \mathcal{C}_{2,-p^2}^{\rm improved}[m^2,J] + \sum_{k=4,6,\dots} \int_0^1 dp\,h_k(p) \mathcal{X}_{k,-p^2}[m^2,J] \geq 0 \nn\\
&\quad \forall m\geq 1,\ J=0,2,4\ldots
\\
\mbox{then: }&\int_0^1 dp\,f(p)\left[\frac{1}{p^2} +2g_2 +g_3p^2\right] \geq 0\,,
\label{eq:fulllinearprogramagain}
\end{align}
We have set $M=1$ and $8\pi G=1$; the dependence on these quantities can be restored using dimensional analysis and homogeneity.

We are free to choose any objective function and normalization condition on the functions $f(p)$ and $h_k(p)$. The solution of the resulting optimization problem then provides a valid inequality on EFT data of the form (\ref{eq:fulllinearprogramagain}). The full allowed region in the $(g_2,g_3)$ plane is the intersection of the allowed regions for all such inequalities. For example, to obtain the plots in Figure~\ref{fig:allowedregionsdifferentD}, we maximized the distance from a chosen point $(g_{2,0},g_{3,0})$ along rays of constant angle $\theta$ in the $(g_2,g_3)$ plane. We chose $(g_{2,0},g_{3,0})$ near the tip of the expected allowed region (known from earlier experimentation) and scanned over angles $\theta\in \{0,\frac{\pi}{20},\dots,\frac{39\pi}{20}\}$.

We expand $f(p)$ and $h_k(p)$ in pure powers of $p$,
\begin{align}
f(p) = \sum_n a_n p^n,\qquad
h_k(p) = \sum_{i=0}^{i_k} b_{k,i} p^i.
\end{align}
For each $J$, the integrals (\ref{eq:thepositivityconditions}) against pure powers of $p$ can be computed analytically in terms of ${}_2F_1$ hypergeometric functions, for example
\begin{align}
&\int_0^1 dp\, p^n \mathcal{C}_{2,-p^2}^\mathrm{improved}[m^2,2] \nn\\
&=
-\frac{4 (D-1) \, _2F_1\left(1,\frac{n+1}{2};\frac{n+3}{2};-\frac{1}{m^2}\right)}{(D-2) m^4 (n+1)}+\frac{2 (3 D-4)}{(D-2) m^4 (n+1)}-\frac{3 (3 D-2)}{(D-2) m^6 (n+3)}.
\label{eq:examplefunctionofm}
\end{align}
Parametrizing $m^2 = \frac{1}{1-x}$, we would ideally like to impose (\ref{eq:thepositivityconditions}) for all $J=0,2,\dots$ and $x\in[0,1)$. In practice, we must restrict $J\in \{0,2,\dots,J_\mathrm{max}\}$ and discretize $x$.\footnote{Alternatively, it would be interesting to find an approximation for functions like (\ref{eq:examplefunctionofm}) in terms of a positive function of $x$ times a polynomial. This would allow us to rewrite positivity constraints in terms of positive semidefinite matrices and apply semidefinite programming, as done for CFT four-point functions \cite{Poland:2011ey,Kos:2014bka}.} Our initial discretization is
\begin{align}
\label{eq:discretizedx}
x\in \{0,\de_x,2\de_x,\dots,\lceil \tfrac{1}{\de_x}-1 \rceil \de_x\}
\end{align}
for some a small parameter $\de_x$, listed below in Table~\ref{tab:parameters}.
Discretizing $x$ weakens the inequalities, potentially resulting in incorrect bounds. However, we can effectively remove this problem by adaptively refining the discretization as described in Section~\ref{sec:adaptiverefinement}.

\subsection{Impact parameter space inequalities}
\label{sec:impactparamineqs}

Restricting $J\leq J_\mathrm{max}$ weakens the inequalities as well. To reduce the dependence of the resulting bounds on $J_\mathrm{max}$, it is useful to explicitly include inequality constraints from the scaling limit $m\to \oo$ with fixed impact parameter $b=\frac{2J}{m}$:
\begin{align}
\Gamma\big(\tfrac{D-2}{2}\big)
 \int_0^1 dp f(p)\ \frac{J_{\frac{D-4}{2}}(bp)}{(bp/2)^{\frac{D-4}{2}}} &\geq 0 \qquad \forall b \geq 0.
\end{align}
(The null constraints $\mathcal{X}_4,\mathcal{X}_6,\dots$ are subleading in this limit, so the functions $h_k(p)$ don't enter this condition.) Again, the integral against a pure power of $p$ can be done analytically:
\begin{align}
\Gamma\big(\tfrac{D-2}{2}\big)
 \int_0^1 dp\, p^n\ \frac{J_{\frac{D-4}{2}}(bp)}{(bp/2)^{\frac{D-4}{2}}}
 &=
\frac{\, _1F_2\left(\frac{n+1}{2};\frac{D-2}{2},\frac{n+3}{2};-\frac{b^2}{4}\right)}{n+1}.
\end{align}
The resulting functions have an oscillatory and non-oscillatory part at large $b$:\footnote{The full expansion of ${}_1F_2$ hypergeometric functions around infinity can be found in \cite{NIST:DLMF}.}
\begin{align}
\label{eq:the1f2s}
\frac{\, _1F_2\left(\frac{n+1}{2};\frac{D-2}{2},\frac{n+3}{2};-\frac{b^2}{4}\right)}{n+1}
&\sim
\frac{2^n \Gamma (\frac{D-2}{2}) \Gamma (\frac{n+1}{2})}{\Gamma (\frac{D-n-3}{2})}\frac{1}{b^{n+1}} +\frac{2^{\frac{D-3}{2}}  \Gamma (\frac{D-2}{2}) 
}{\sqrt{\pi }}\frac{\cos \p{b-\frac{\pi  (D-1)}{4}}}{b^{\frac{D-1}{2}}} + \dots
\end{align}
In order for linear combinations of such functions to be positive at large $b$, we must either include at least one $n$ such that $n\leq \frac{D-3}{2}$, or take linear combinations of functions that cancel the leading oscillatory term at large $b$. We return to this statement below.

To efficiently impose positivity at large $b$, we use the following trick. After writing $f(p)$ as a sum of pure powers, we have
\begin{align}
\label{eq:thingthatispositive}
\Gamma\big(\tfrac{D-2}{2}\big)
 \int_0^1 dp f(p)\ \frac{J_{\frac{D-4}{2}}(bp)}{(bp/2)^{\frac{D-4}{2}}} &= A(b) + B(b) \cos \p{b-\frac{\pi  (D-1)}{4}} + C(b) \sin \p{b-\frac{\pi  (D-1)}{4}},
\end{align}
where the functions $A(b), B(b), C(b)$ have well-behaved asymptotic expansions in inverse powers of $b$. Let us now write
\begin{align}
A+B\cos\phi+ C\sin\phi &= 
\begin{pmatrix}
\cos \frac{\phi}{2} &
\sin \frac{\phi}{2}
\end{pmatrix}
\begin{pmatrix}
A+B & C \\
C & A-B
\end{pmatrix}
\begin{pmatrix}
\cos \frac{\phi}{2} \\
\sin \frac{\phi}{2}
\end{pmatrix}.
\end{align}
We can thus replace positivity of (\ref{eq:thingthatispositive}) with the stronger condition
\begin{align}
\begin{pmatrix}
A(b)+B(b) & C(b) \\
C(b) & A(b)-B(b)
\end{pmatrix} \succeq 0,
\label{eq:strongercondition}
\end{align}
where ``$M\succeq 0$'' means $M$ is positive semidefinite. Because (\ref{eq:strongercondition}) implies (\ref{eq:thingthatispositive}), this replacement is rigorous, but may result in sub-optimal bounds. However, at large $b$, the $\cos(b-\frac{\pi(D-1)}{4})$ and $\sin(b-\frac{\pi(D-1)}{4})$ terms in (\ref{eq:thingthatispositive}) are rapidly oscillating so that (\ref{eq:strongercondition}) is a good approximation to the original positivity condition. In (\ref{eq:strongercondition}), we can now expand $A(b), B(b), C(b)$ in inverse powers of $b$, and truncate the expansions. If $n\equiv \frac{D-3}{2} \mod 1$, after pulling out a positive factor, we obtain a matrix polynomial inequality of $b$, which can be used in {\tt SDPB}.\footnote{Alternatively, if $n\equiv \frac{D-3}{2}+\frac p q \mod 1$, we can perform a change of variables $b\to b'^q$ to again obtain a matrix polynomial of $b'$. However, this step multiplies the degree of the resulting polynomial by $q$, which can result in a performance hit in {\tt SDPB}.}  

These considerations suggest that we should expand $f(p)$ in the functions $p^2,p^3,\dots$ in odd $D$ and the functions $p^{3/2},p^{5/2},\dots$ in even $D$. This works well except in $D=5$, since $2>\frac{5-3}{2}$, so generic linear combinations of (\ref{eq:the1f2s}) cannot be positive at large $b$. One solution in $D=5$ is to cancel the leading oscillatory term at large $b$, which can be done by using differences $p^3-p^2,p^4-p^2,\dots$. This leads to the basis functions shown in Table~\ref{tab:basisfunctionsforf}.

We encode positivity using the $2\x 2$ matrix (\ref{eq:strongercondition}) for $b\geq B$, where $B$ is some cutoff. We keep $m_\mathrm{max}$ subleading terms in the expansion of $B(b)$ and $C(b)$ at large $b$, where $m_\mathrm{max}$ is listed below. For smaller $b\leq B$, we impose positivity at discretized impact parameters:
\begin{align}
b \in \{\e_b,\e_b+\de_b,\dots,\e_b + \lceil\tfrac{B-\e_b}{\de_b}-1\rceil \de_b\},
\end{align}
where $\e_b,\de_b$ are small parameters listed below in Table~\ref{tab:parameters}. Like in the fixed-$J$ case, we adaptively refine our discretization of $b$, as described in Section~\ref{sec:adaptiverefinement}. 

\subsection{Outer approximation/adaptive refinement}
\label{sec:adaptiverefinement}

Our problem has several constraints that depend on a continuous parameter. For example, we would like to impose positivity of (\ref{eq:thepositivityconditions}) for all $x\in[0,1)$ and $J=0,2,4,\dots$, where $m=\frac{1}{1-x}$. By only imposing positivity at a discrete set of $x$ as in (\ref{eq:discretizedx}), we run the risk that the solver could return a solution that is negative between two discretized values. In fact, this almost always happens, since the solution to any optimization problem involves some set of saturated inequalities. If the left-hand side of  (\ref{eq:thepositivityconditions}) is zero at some value of $x$, it will generically be negative on one side of that zero. Typically, the solver returns a solution that vanishes at pairs of neighboring discrete values of $x$, and is negative between them.

\begin{figure}
\centering
\includegraphics[width=0.7\textwidth]{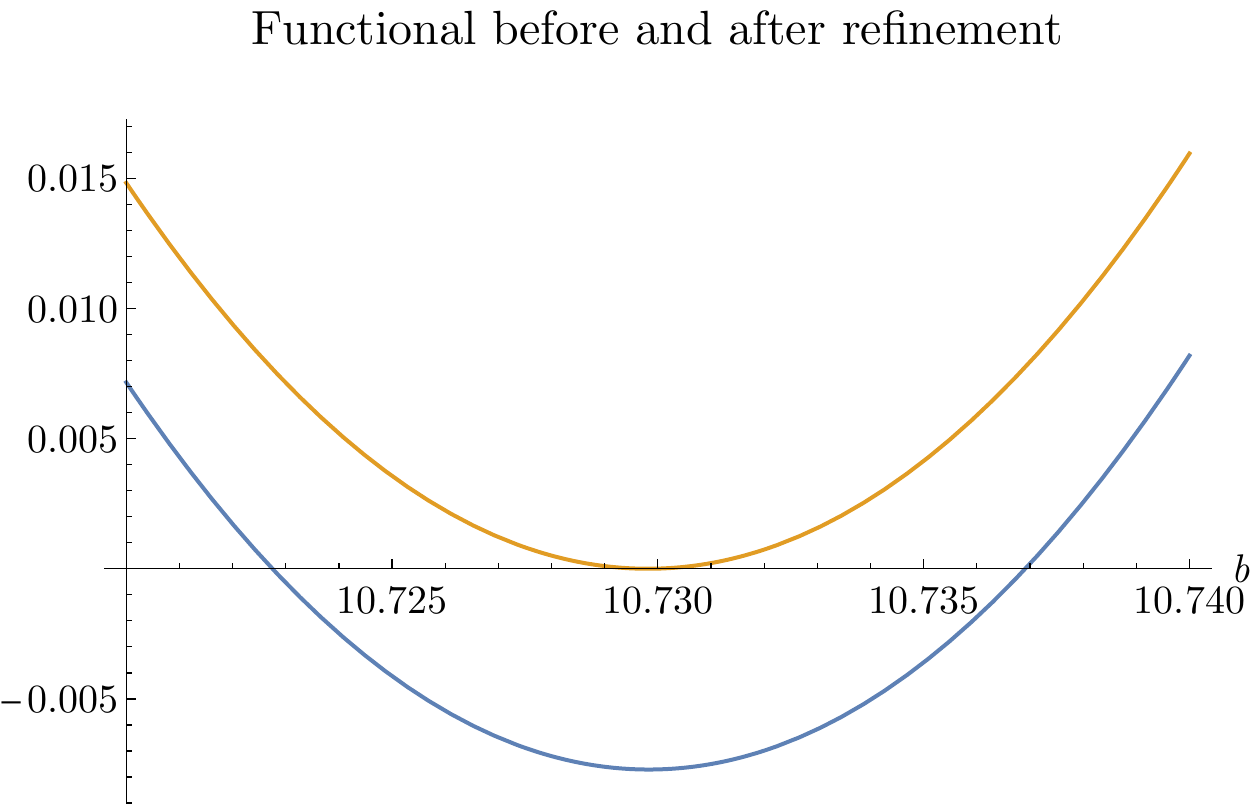}
\caption{
A functional returned by {\tt SDPB} before refinement (blue) and after 3 steps of refining the discretization and re-running {\tt SDPB} (orange). We plot the action of the functional in impact parameter space, zoomed in near a negative region after the initial run of {\tt SDPB}. Our initial discretization used step size $\de_b = 1/32$. The functional indeed has a negative region of roughly size $\de_b$, with minimum value $-0.0077$ in this region. After the refinement steps, the negative region is $b\in (10.7298112,10.7298134)$, and the minimum value of the functional is $-1.83\x 10^{-10}$.
\label{fig:refinement}
}
\end{figure}

We mitigate this problem by adaptively refining the discretization. We begin with an initial discretization of $x$ and $b$ and run the solver. The resulting functional will be negative between pairs of points in the initial discretization. We identify these negative regions and add new positivity constraints in a finer-spaced grid covering the negative regions. Specifically, suppose that the functional dips negative between a pair $(x_1,x_2)$ of discretized $x$. Let the minimum of the functional between $x_1$ and $x_2$ (which we estimate from a quadratic approximation) be $x_*$. We add new positivity constraints at the locations
\be
\{ x_* - N s,\, x_* - (N-1) s,\,\dots,\,x_*,\,\dots,\,x_*+(N-1) s,\, x_* + N s \},
\ee
where $s=|x_2-x_1|/N$ with e.g.\ $N=10$. Running the solver again with the new constraints included, the new solution will typically have a saturated inequality near $x_*$ with a negative region reduced in size by a factor of $N$. We repeat this procedure until the negative regions are extremely small (e.g.\ of size $10^{-6}$), and it is clear that the solution is converging to a nonnegative functional. The solution after refinement is usually quite close to the original solution. We give an example of a negative region in an initial solution and the result after refining the discretization in Figure~\ref{fig:refinement}.

An important benefit of this method is that the bounds become essentially independent of the parameters $\e_x,\de_x,\e_b,\de_b$ describing the initial discretization, provided they are sufficiently small.

Abstractly, the space of allowed functions $f(p), h_k(p)$ is a convex region carved out by an infinite number of inequalities. By discretizing $x$, we obtain an ``outer approximation'' of this region in terms of a finite number of inequalities. (``Outer'' because our approximate region is bigger than the true region.) By refining our discretization, we obtain a more accurate outer approximation. An efficient implementation of this method should include a way of hot-starting from the previous solution after each refinement step. We have not implemented this --- instead we simply run {\tt SDPB} from scratch after each refinement. Typically only a few refinement steps are needed, so this does not give a huge performance hit.

\subsection{Choices of parameters and numerical results}

\begin{table}
\centering
\begin{tabular}{c|c|c}
 & Figure~\ref{fig:allowedregionsdifferentD} & Figure~\ref{fig:extremalspectrum} \\
\hline
functionals & 
\begin{tabular}{@{}l@{}}
$\{p^n \mathcal{C}_2^\mathrm{improved}\,|\,n\leq 7.5\}$ 
 \\
 $\cup\, \{ p^i \mathcal{X}_4\,|\,i = 0,1,2,3,4,5 \}$
 \\
 $\cup\, \{ p^i \mathcal{X}_6\,|\,i = 0,1,2,3 \}$
\end{tabular}
&
\begin{tabular}{@{}l@{}}
$\{p^n \mathcal{C}_2^\mathrm{improved}\,|\,n=1.5,2.5,\dots,27.5\}$
\\
 $\cup\, \{ p^i \mathcal{X}_4\,|\,i = 0,1,\dots,15 \}$
\\
 $\cup\, \{ p^i \mathcal{X}_6\,|\,i = 0,1,\dots,13 \}$
\\
$\cup\, \dots$
\\
$\cup\, \{ p^i \mathcal{X}_{14}\,|\,i = 0,1,\dots,5 \}$
\end{tabular}
 \\
\hline
$J_\mathrm{max}$ & 42 & 150 \\
$\de_x$ & 1/400 & 1/800  \\
$\e_b$ & 1/250 & 1/250 \\
$\de_b$ & 1/32 & 1/100 \\
$B$ & 40 & 80 \\
$m_\mathrm{max}$ & 2 & 6 \\
\hline
\begin{tabular}{@{}c@{}}
 non-default \\
{\tt SDPB} parameters
 \end{tabular}
 & {\tt ----precision=768} &
\begin{tabular}{@{}c@{}}
 {\tt ----precision=840} \\
 {\tt ----dualityGapThreshold=1e-80 } \\
 {\tt ----primalErrorThreshold=1e-80 } \\
 {\tt ----dualErrorThreshold=1e-80 }
\end{tabular}
\end{tabular}
\caption{
Choices of parameters for the computations in this work. These parameters were chosen experimentally --- we have not attempted to optimize them. For the computation in Figure~\ref{fig:allowedregionsdifferentD}, we verified that further increasing $J_\mathrm{max}$ changes the bounds by less than the fraction $10^{-4}$. Because of the refinement procedure described in Section~\ref{sec:adaptiverefinement}, the bounds are essentially independent of $\de_x, \e_b, \de_b$ in all cases. For Figure~\ref{fig:allowedregionsdifferentD}, the precise linear combinations of $\mathcal{C}_2^\mathrm{improved}$ are $D$-dependent, and listed in Table~\ref{tab:basisfunctionsforf}. For the computation in Figure~\ref{fig:extremalspectrum}, we supplemented the initial discretization of $x$ with additional constraints at $x\in \{1-4/J,1-4/J + 1/(50 J),\dots,1-1/(50 J)\}$ for $J=36,\dots,150$. Finally, for the extremal functional computation in Figure~\ref{fig:extremalspectrum}, we project onto the subspace of functionals with vanishing contribution of $g_3$. This reduces the space of functionals from $93$ dimensions to $92$ dimensions. Tables of functionals were produced using Mathematica with 300 decimal digits of precision.
\label{tab:parameters}
}
\end{table}

Our implementation of the linear program (\ref{eq:fulllinearprogram}) involves several parameters. In Table~\ref{tab:parameters}, we list the parameters used for the computations in this work. The inequalities plotted in Figure~\ref{fig:allowedregionsdifferentD} (in units where $8\pi G=1$ and $M=1$) are given in Table~\ref{eq:inequalitiesforfigure1}.

\begin{table}[h]
\begin{align*}
\begin{array}{l|l}
D=5 & 
\begin{array}{ll}
g_ 2-\frac{g_ 3}{3}+60.3086\geq 0 & \\
\land\ \ g_2+0.0647867 g_ 3+9.64034\geq 0 &\ \land\ \ g_2+0.0750150 g_ 3+8.40592\geq 0 \\
\land\ \ g_2+0.0779037 g_ 3+8.09643\geq 0 &\ \land\ \ g_2+0.0802745 g_ 3+7.86165\geq 0 \\
\land\ \ g_2+0.0823523 g_ 3+7.66918\geq 0 &\ \land\ \ g_2+0.0842715 g_ 3+7.50180\geq 0 \\
\land\ \ g_2+0.0861338 g_ 3+7.34848\geq 0 &\ \land\ \ g_2+0.0879254 g_ 3+7.20921\geq 0 \\
\land\ \ g_2+0.0898265 g_ 3+7.07029\geq 0 &\ \land\ \ g_2+0.0919903 g_ 3+6.92264\geq 0 \\
\land\ \ g_2+0.0946037 g_ 3+6.75833\geq 0 &\ \land\ \ g_2+0.0980028 g_ 3+6.56596\geq 0 \\
\land\ \ g_2+0.112285 g_ 3+5.90160\geq 0 &\ \land\ \ g_2+0.112297 g_ 3+5.90112\geq 0 \\
\land\ \ g_2+0.112318 g_ 3+5.90066\geq 0 &\ \land\ \ g_2+0.112362 g_ 3+5.90107\geq 0
\end{array}
\\
\hline
D=6 & 
\begin{array}{ll}
g_ 2-\frac{g_ 3}{3}+28.0546\geq 0 & \\
\land\ \ g_2+0.0940931 g_ 3+4.55838\geq 0 &\ \land\ \ g_2+0.101479 g_ 3+4.22726\geq 0 \\
\land\ \ g_2+0.107010 g_ 3+4.00645\geq 0 &\ \land\ \ g_2+0.110538 g_ 3+3.87630\geq 0 \\
\land\ \ g_2+0.113142 g_ 3+3.78572\geq 0 &\ \land\ \ g_2+0.115195 g_ 3+3.71765\geq 0 \\
\land\ \ g_2+0.116954 g_ 3+3.66176\geq 0 &\ \land\ \ g_2+0.118572 g_ 3+3.61228\geq 0 \\
\land\ \ g_2+0.120160 g_ 3+3.56553\geq 0 &\ \land\ \ g_2+0.121815 g_ 3+3.51866\geq 0
\end{array}
\\
\hline
D=7 & 
\begin{array}{ll}
g_ 2-\frac{g_ 3}{3}+18.0717\geq 0 & \\
\land\ \ g_2+0.107754 g_ 3+3.30601\geq 0 &\ \land\ \ g_2+0.114745 g_ 3+3.11426\geq 0 \\
\land\ \ g_2+0.119761 g_ 3+2.98759\geq 0 &\ \land\ \ g_2+0.124098 g_ 3+2.88546\geq 0 \\
\land\ \ g_2+0.127812 g_ 3+2.80238\geq 0 &\ \land\ \ g_2+0.129850 g_ 3+2.75853\geq 0
\end{array}
\\
\hline
D=8 & 
\begin{array}{ll}
g_ 2-\frac{g_ 3}{3}+13.7186\geq 0 & \\
\land\ \ g_2+0.110239 g_ 3+2.92120\geq 0 &\ \land\ \ g_2+0.113486 g_ 3+2.84526\geq 0 \\
\land\ \ g_2+0.116510 g_ 3+2.77802\geq 0 &\ \land\ \ g_2+0.119462 g_ 3+2.71525\geq 0 \\
\land\ \ g_2+0.122467 g_ 3+2.65393\geq 0 &\ \land\ \ g_2+0.125690 g_ 3+2.59077\geq 0 \\
\land\ \ g_2+0.129207 g_ 3+2.52462\geq 0 &\ \land\ \ g_2+0.133224 g_ 3+2.45244\geq 0 \\
\land\ \ g_2+0.134841 g_ 3+2.42430\geq 0 &\ \land\ \ g_2+0.134849 g_ 3+2.42422\geq 0 \\
\land\ \ g_2+0.134858 g_ 3+2.42416\geq 0 &\ \land\ \ g_2+0.134874 g_ 3+2.42429\geq 0
\end{array}
\\
\hline
D=9 & 
\begin{array}{ll}
g_ 2-\frac{g_ 3}{3}+11.1250\geq 0 & \\
\land\ \ g_2+0.134185 g_ 3+2.24546\geq 0 &\ \land\ \ g_2+0.138540 g_ 3+2.17837\geq 0 \\
\land\ \ g_2+0.138548 g_ 3+2.17837\geq 0 &\ \land\ \ g_2+0.138567 g_ 3+2.17856\geq 0
\end{array}
\\
\hline
D=10 & 
\begin{array}{ll}
g_ 2-\frac{g_ 3}{3}+9.5208\geq 0 & \\
\land\ \ g_2+0.138871 g_ 3+2.05513\geq 0 &\ \land\ \ g_2+0.14150 g_ 3+2.0196\geq 0
\end{array}
\\
\hline
D=11 & 
\begin{array}{ll}
g_ 2-\frac{g_ 3}{3}+8.4687\geq 0 & \\
\land\ \ g_2+0.143720 g_ 3+1.91178\geq 0 &\ \land\ \ g_2+0.143729 g_ 3+1.91169\geq 0 \\
\land\ \ g_2+0.143733 g_ 3+1.91166\geq 0 &\ \land\ \ g_2+0.143738 g_ 3+1.91165\geq 0 \\
\land\ \ g_2+0.143746 g_ 3+1.91166\geq 0 &\ \land\ \ g_2+0.143772 g_ 3+1.91187\geq 0
\end{array}
\\
\hline
D=12 & 
\begin{array}{ll}
g_ 2-\frac{g_ 3}{3}+7.9016\geq 0 & \\
\land\ \ g_2+0.144433 g_ 3+1.86671\geq 0 &\ \land\ \ g_2+0.145825 g_ 3+1.85029\geq 0
\end{array}
\end{array}
\end{align*}
\caption{
The inequalities plotted in Figure~\ref{fig:allowedregionsdifferentD} (in units where $8\pi G=1$ and $M=1$).
\label{eq:inequalitiesforfigure1}
}
\end{table}
\clearpage

\section{Bounds using an extended range of $u$}
\label{app:extendedu}

When $u=-M^2$, the $t$-channel cut merges with the origin in the $s$-plane. As noted in section~\ref{sec:secondstrategy}, this may invalidate the $\mathcal{C}_{2,u}^\mathrm{improved}$ sum rule in general for $u\leq-M^2$. However, for meromorphic amplitudes, i.e.\ amplitudes where the $t$-channel ``cut'' is simply a collection of simple poles, there is no obvious problem with taking $u\leq-M^2$. Indeed, the low-energy contribution to the improved sum rules $\mathcal{C}_{k,u}^\mathrm{improved}$ and $\mathcal{X}_{k,u}$ stays finite for general $u<0$. This is unlike the situation for the unimproved sum rule $\mathcal{C}_{k,u}$, the left-hand side of which is an infinite sum of contact contributions with a finite radius of convergence. Similarly, the terms on the right-hand side of $\mathcal{C}_{k,u}$ have a pole at $u=-m^2$, but this pole is removed in the improved sum rules. We have checked in examples coming from string theory, such as \eqref{eq:MString}, that the improved sum rules continue to hold in the complete range $u\in (-\oo,0)$. In general, we expect the improved sum rules to hold in the complete range if the exchanged spectrum contains only finitely many states below any given mass, as is the case in string theory. This is because any finite number of states can not affect the convergence of the improved sum rules and by removing all states with $m^2<-u$, we effectively find ourselves in the situation $u\in(-M^2,0)$ where convergence is guaranteed.

\begin{figure}
\centering
\includegraphics[width=0.8\textwidth]{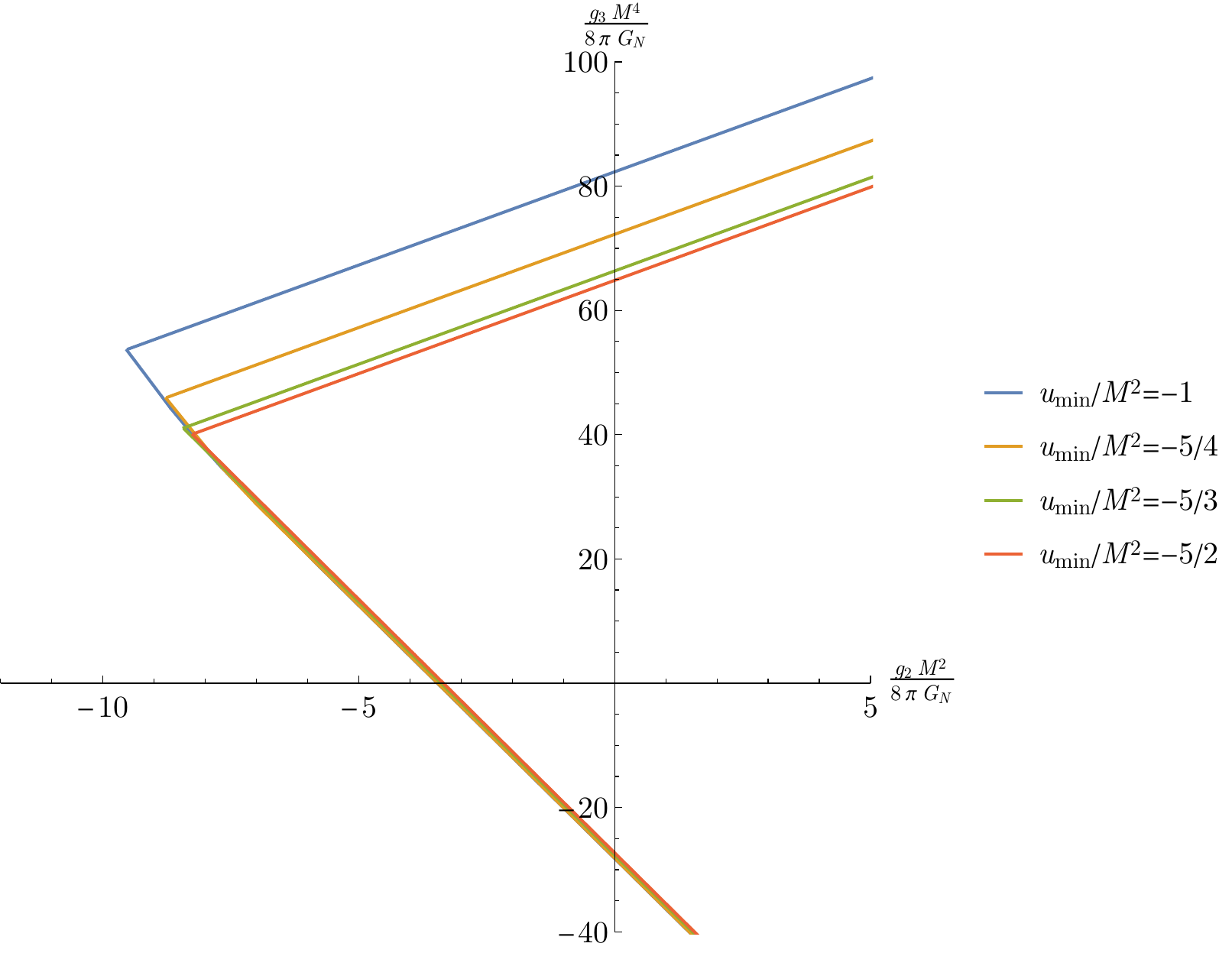}
\caption{
Bounds on $g_2,g_3$ in $D=6$, computed using wavepackets with support in $u\in [u_\mathrm{min},0]$ for $u_\mathrm{min}/M^2=-1,-5/4,-5/3,-5/2$. As $u_\mathrm{min}$ gets more negative, the kink slides along the lower bound on $g_3$. These bounds were computed using a 54-dimensional space of functionals spanned by $\{p^n\mathcal{C}_{2,-p^2}^{\mathrm{improved}}\,|\,n=1.5,\dots,19.5\}\cup \{ p^n \mathcal{X}_k\,|\,(n,k)\in \{(0,4),\dots,(11,4)\}\cup\{(0,6),\dots,(9,6)\} \cup \{(0,8),\dots,(7,8)\} \cup \{(0,10),\dots,(5,10)\}\}$.
\label{fig:extendedrangebounds}
}
\end{figure}

\begin{figure}
\centering
\includegraphics[width=\textwidth]{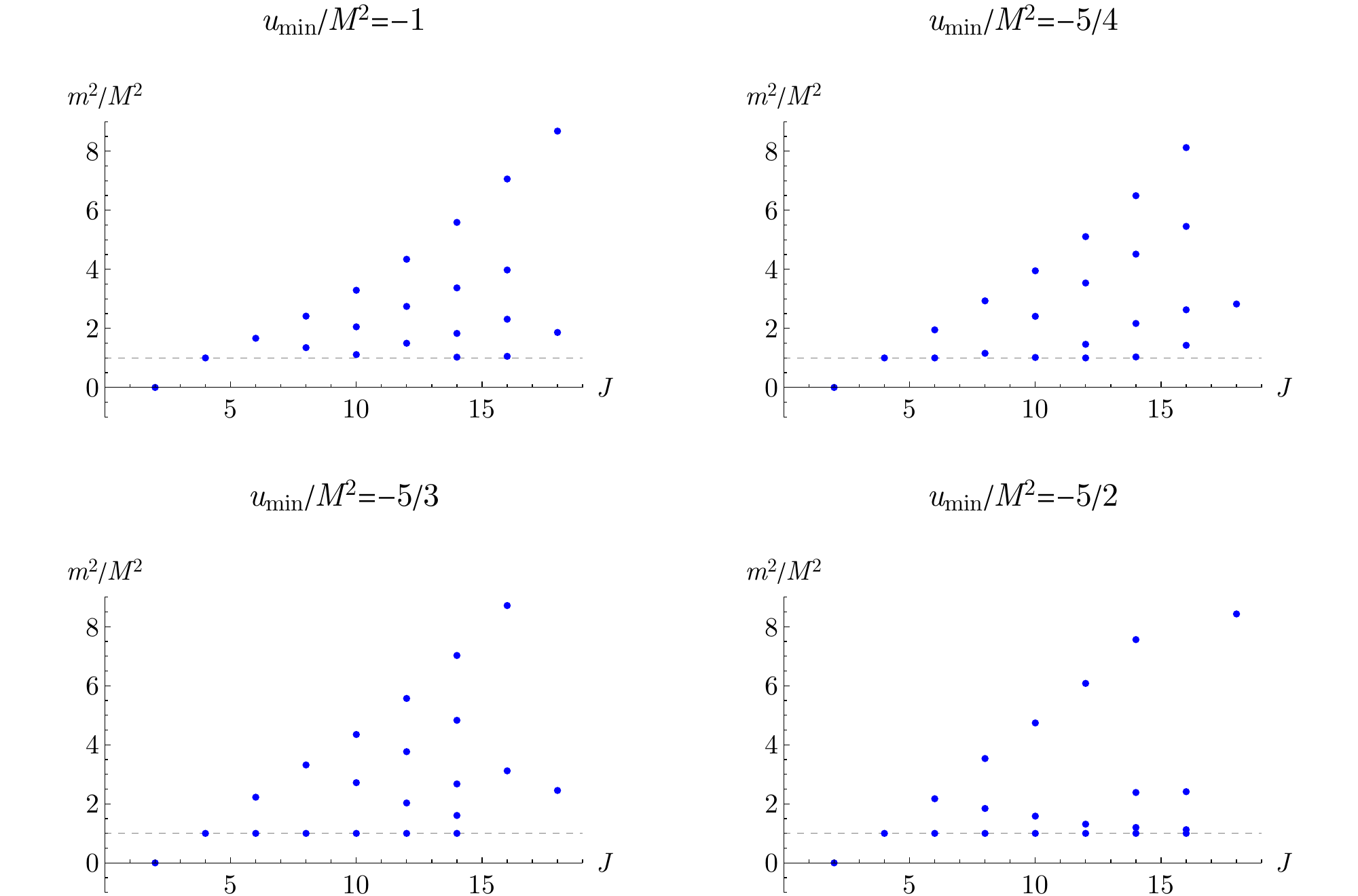}
\caption{
Extremal spectra for the bound that minimizes $g_2$ in $D=6$, computed using different wavefunctions with different ranges $u\in [u_\mathrm{min},0]$. As $u_\mathrm{min}$ gets more negative, the spectrum appears to simplify, with  approximately linear Regge trajectories at $m^2\sim M^2(J-2)/2$ and $m^2=M^2$.
\label{fig:extendedrangespectra}
}
\end{figure}

Thus, it is interesting to consider how the bounds change when we use wavefunctions with larger support in $u$. We leave a more complete exploration of this idea to future work. In this appendix, we explore some example bounds obtained with a larger (but still compact) range of $u$. In figure~\ref{fig:extendedrangebounds}, we show bounds on $g_2,g_3$ obtained using wavefunctions with support on $u\in [-yM^2,0]$ for various $y$. As $y$ increases, the minimal $g_2$ ``kink'' slides to the right. Interestingly, the corresponding extremal spectra appear to simplify, see figure~\ref{fig:extendedrangespectra}. For larger $y$, the spectrum may be converging toward a single linear Regge trajectory $m^2=M^2(J-2)/2$, with possibly another trajectory at $m^2=M^2$. It is interesting to ask whether a solution to crossing symmetry with this structure can exist.

\clearpage

\bibliographystyle{./aux/ytphys}
\bibliography{./aux/refs}

\providecommand{\href}[2]{#2}\begingroup\raggedright\begin{thebibliography}{10}

\bibitem{Ooguri:2006in}
H.~Ooguri and C.~Vafa, ``{On the Geometry of the String Landscape and the
  Swampland},'' \href{http://dx.doi.org/10.1016/j.nuclphysb.2006.10.033}{{\em
  Nucl. Phys. B} {\bfseries 766} (2007) 21--33},
  \href{http://arxiv.org/abs/hep-th/0605264}{{\ttfamily arXiv:hep-th/0605264}}.

\bibitem{Brennan:2017rbf}
T.~D. Brennan, F.~Carta, and C.~Vafa, ``{The String Landscape, the Swampland,
  and the Missing Corner},'' \href{http://dx.doi.org/10.22323/1.305.0015}{{\em
  PoS} {\bfseries TASI2017} (2017) 015},
  \href{http://arxiv.org/abs/1711.00864}{{\ttfamily arXiv:1711.00864
  [hep-th]}}.

\bibitem{Palti:2019pca}
E.~Palti, ``{The Swampland: Introduction and Review},''
  \href{http://dx.doi.org/10.1002/prop.201900037}{{\em Fortsch. Phys.}
  {\bfseries 67} no.~6, (2019) 1900037},
  \href{http://arxiv.org/abs/1903.06239}{{\ttfamily arXiv:1903.06239
  [hep-th]}}.

\bibitem{vanBeest:2021lhn}
M.~van Beest, J.~Calder\'on-Infante, D.~Mirfendereski, and I.~Valenzuela,
  ``{Lectures on the Swampland Program in String Compactifications},''
  \href{http://arxiv.org/abs/2102.01111}{{\ttfamily arXiv:2102.01111
  [hep-th]}}.

\bibitem{ArkaniHamed:2006dz}
N.~Arkani-Hamed, L.~Motl, A.~Nicolis, and C.~Vafa, ``{The String landscape,
  black holes and gravity as the weakest force},''
  \href{http://dx.doi.org/10.1088/1126-6708/2007/06/060}{{\em JHEP} {\bfseries
  06} (2007) 060}, \href{http://arxiv.org/abs/hep-th/0601001}{{\ttfamily
  arXiv:hep-th/0601001}}.

\bibitem{Pham:1985cr}
T.~N. Pham and T.~N. Truong, ``{Evaluation of the Derivative Quartic Terms of
  the Meson Chiral Lagrangian From Forward Dispersion Relation},''
  \href{http://dx.doi.org/10.1103/PhysRevD.31.3027}{{\em Phys. Rev. D}
  {\bfseries 31} (1985) 3027}.

\bibitem{Adams:2006sv}
A.~Adams, N.~Arkani-Hamed, S.~Dubovsky, A.~Nicolis, and R.~Rattazzi,
  ``{Causality, analyticity and an IR obstruction to UV completion},''
  \href{http://dx.doi.org/10.1088/1126-6708/2006/10/014}{{\em JHEP} {\bfseries
  10} (2006) 014}, \href{http://arxiv.org/abs/hep-th/0602178}{{\ttfamily
  arXiv:hep-th/0602178}}.

\bibitem{Arkani-Hamed:2020blm}
N.~Arkani-Hamed, T.-C. Huang, and Y.-T. Huang, ``{The EFT-Hedron},''
  \href{http://arxiv.org/abs/2012.15849}{{\ttfamily arXiv:2012.15849
  [hep-th]}}.

\bibitem{Bellazzini:2020cot}
B.~Bellazzini, J.~Elias~Mir\'o, R.~Rattazzi, M.~Riembau, and F.~Riva,
  ``{Positive Moments for Scattering Amplitudes},''
  \href{http://arxiv.org/abs/2011.00037}{{\ttfamily arXiv:2011.00037
  [hep-th]}}.

\bibitem{Tolley:2020gtv}
A.~J. Tolley, Z.-Y. Wang, and S.-Y. Zhou, ``{New positivity bounds from full
  crossing symmetry},'' \href{http://arxiv.org/abs/2011.02400}{{\ttfamily
  arXiv:2011.02400 [hep-th]}}.

\bibitem{Caron-Huot:2020cmc}
S.~Caron-Huot and V.~Van~Duong, ``{Extremal Effective Field Theories},''
  \href{http://arxiv.org/abs/2011.02957}{{\ttfamily arXiv:2011.02957
  [hep-th]}}.

\bibitem{Sinha:2020win}
A.~Sinha and A.~Zahed, ``{Crossing Symmetric Dispersion Relations in QFTs},''
  \href{http://dx.doi.org/10.1103/PhysRevLett.126.181601}{{\em Phys. Rev.
  Lett.} {\bfseries 126} no.~18, (2021) 181601},
  \href{http://arxiv.org/abs/2012.04877}{{\ttfamily arXiv:2012.04877
  [hep-th]}}.

\bibitem{Froissart:1961ux}
M.~Froissart, ``{Asymptotic behavior and subtractions in the Mandelstam
  representation},'' \href{http://dx.doi.org/10.1103/PhysRev.123.1053}{{\em
  Phys. Rev.} {\bfseries 123} (1961) 1053--1057}.

\bibitem{Martin:1962rt}
A.~Martin, ``{Unitarity and high-energy behavior of scattering amplitudes},''
  \href{http://dx.doi.org/10.1103/PhysRev.129.1432}{{\em Phys. Rev.} {\bfseries
  129} (1963) 1432--1436}.

\bibitem{PhysRev.95.1612}
M.~Gell-Mann, M.~L. Goldberger, and W.~E. Thirring, ``Use of causality
  conditions in quantum theory,''
  \href{https://link.aps.org/doi/10.1103/PhysRev.95.1612}{{\em Phys. Rev.}
  {\bfseries 95} (Sep, 1954) 1612--1627}.

\bibitem{PhysRev.99.979}
M.~L. Goldberger, ``Causality conditions and dispersion relations. i. boson
  fields,'' \href{https://link.aps.org/doi/10.1103/PhysRev.99.979}{{\em Phys.
  Rev.} {\bfseries 99} (Aug, 1955) 979--985}.

\bibitem{Camanho:2014apa}
X.~O. Camanho, J.~D. Edelstein, J.~Maldacena, and A.~Zhiboedov, ``{Causality
  Constraints on Corrections to the Graviton Three-Point Coupling},''
  \href{http://dx.doi.org/10.1007/JHEP02(2016)020}{{\em JHEP} {\bfseries 02}
  (2016) 020}, \href{http://arxiv.org/abs/1407.5597}{{\ttfamily arXiv:1407.5597
  [hep-th]}}.

\bibitem{Chowdhury:2019kaq}
S.~D. Chowdhury, A.~Gadde, T.~Gopalka, I.~Halder, L.~Janagal, and S.~Minwalla,
  ``{Classifying and constraining local four photon and four graviton
  S-matrices},'' \href{http://dx.doi.org/10.1007/JHEP02(2020)114}{{\em JHEP}
  {\bfseries 02} (2020) 114}, \href{http://arxiv.org/abs/1910.14392}{{\ttfamily
  arXiv:1910.14392 [hep-th]}}.

\bibitem{Chandorkar:2021viw}
D.~Chandorkar, S.~D. Chowdhury, S.~Kundu, and S.~Minwalla, ``{Bounds on Regge
  growth of flat space scattering from bounds on chaos},''
  \href{http://arxiv.org/abs/2102.03122}{{\ttfamily arXiv:2102.03122
  [hep-th]}}.

\bibitem{Susskind:1998vk}
L.~Susskind, ``{Holography in the flat space limit},''
  \href{http://dx.doi.org/10.1063/1.1301570}{{\em AIP Conf. Proc.} {\bfseries
  493} no.~1, (1999) 98--112},
  \href{http://arxiv.org/abs/hep-th/9901079}{{\ttfamily arXiv:hep-th/9901079}}.

\bibitem{Polchinski:1999ry}
J.~Polchinski, ``{S matrices from AdS space-time},''
  \href{http://arxiv.org/abs/hep-th/9901076}{{\ttfamily arXiv:hep-th/9901076}}.

\bibitem{Penedones:2010ue}
J.~Penedones, ``{Writing CFT correlation functions as AdS scattering
  amplitudes},'' \href{http://dx.doi.org/10.1007/JHEP03(2011)025}{{\em JHEP}
  {\bfseries 03} (2011) 025},
\href{http://arxiv.org/abs/1011.1485}{{\ttfamily arXiv:1011.1485 [hep-th]}}.

\bibitem{Caron-Huot:2017vep}
S.~Caron-Huot, ``{Analyticity in Spin in Conformal Theories},''
  \href{http://dx.doi.org/10.1007/JHEP09(2017)078}{{\em JHEP} {\bfseries 09}
  (2017) 078},
\href{http://arxiv.org/abs/1703.00278}{{\ttfamily arXiv:1703.00278 [hep-th]}}.

\bibitem{Maldacena:2015waa}
J.~Maldacena, S.~H. Shenker, and D.~Stanford, ``{A bound on chaos},''
  \href{http://dx.doi.org/10.1007/JHEP08(2016)106}{{\em JHEP} {\bfseries 08}
  (2016) 106}, \href{http://arxiv.org/abs/1503.01409}{{\ttfamily
  arXiv:1503.01409 [hep-th]}}.

\bibitem{Bellazzini:2019xts}
B.~Bellazzini, M.~Lewandowski, and J.~Serra, ``{Positivity of Amplitudes, Weak
  Gravity Conjecture, and Modified Gravity},''
  \href{http://dx.doi.org/10.1103/PhysRevLett.123.251103}{{\em Phys. Rev.
  Lett.} {\bfseries 123} no.~25, (2019) 251103},
  \href{http://arxiv.org/abs/1902.03250}{{\ttfamily arXiv:1902.03250
  [hep-th]}}.

\bibitem{Tokuda:2020mlf}
J.~Tokuda, K.~Aoki, and S.~Hirano, ``{Gravitational positivity bounds},''
  \href{http://dx.doi.org/10.1007/JHEP11(2020)054}{{\em JHEP} {\bfseries 11}
  (2020) 054}, \href{http://arxiv.org/abs/2007.15009}{{\ttfamily
  arXiv:2007.15009 [hep-th]}}.

\bibitem{Alberte:2020jsk}
L.~Alberte, C.~de~Rham, S.~Jaitly, and A.~J. Tolley, ``{Positivity Bounds and
  the Massless Spin-2 Pole},''
  \href{http://dx.doi.org/10.1103/PhysRevD.102.125023}{{\em Phys. Rev. D}
  {\bfseries 102} (2020) 125023},
  \href{http://arxiv.org/abs/2007.12667}{{\ttfamily arXiv:2007.12667
  [hep-th]}}.

\bibitem{Pajer:2020wnj}
E.~Pajer, D.~Stefanyszyn, and J.~Supel, ``{The Boostless Bootstrap: Amplitudes
  without Lorentz boosts},''
  \href{http://dx.doi.org/10.1007/JHEP12(2020)198}{{\em JHEP} {\bfseries 12}
  (2020) 198}, \href{http://arxiv.org/abs/2007.00027}{{\ttfamily
  arXiv:2007.00027 [hep-th]}}.

\bibitem{Grall:2021xxm}
T.~Grall and S.~Melville, ``{Positivity Bounds without Boosts},''
  \href{http://arxiv.org/abs/2102.05683}{{\ttfamily arXiv:2102.05683
  [hep-th]}}.

\bibitem{Alberte:2020bdz}
L.~Alberte, C.~de~Rham, S.~Jaitly, and A.~J. Tolley, ``{QED positivity
  bounds},'' \href{http://arxiv.org/abs/2012.05798}{{\ttfamily arXiv:2012.05798
  [hep-th]}}.

\bibitem{CHMRSD3}
S.~Caron-Huot, D.~Maz\'{a}\v{c}, L.~Rastelli, and D.~Simmons-Duffin, ``{to
  appear}.''.

\bibitem{Heemskerk:2009pn}
I.~Heemskerk, J.~Penedones, J.~Polchinski, and J.~Sully, ``{Holography from
  Conformal Field Theory},''
  \href{http://dx.doi.org/10.1088/1126-6708/2009/10/079}{{\em JHEP} {\bfseries
  10} (2009) 079}, \href{http://arxiv.org/abs/0907.0151}{{\ttfamily
  arXiv:0907.0151 [hep-th]}}.

\bibitem{Carmi:2019cub}
D.~Carmi and S.~Caron-Huot, ``{A Conformal Dispersion Relation: Correlations
  from Absorption},''
\href{http://arxiv.org/abs/1910.12123}{{\ttfamily arXiv:1910.12123 [hep-th]}}.

\bibitem{Mazac:2019shk}
D.~Maz\'a\v{c}, L.~Rastelli, and X.~Zhou, ``{A Basis of Analytic Functionals
  for CFTs in General Dimension},''
\href{http://arxiv.org/abs/1910.12855}{{\ttfamily arXiv:1910.12855 [hep-th]}}.

\bibitem{Penedones:2019tng}
J.~Penedones, J.~A. Silva, and A.~Zhiboedov, ``{Nonperturbative Mellin
  Amplitudes: Existence, Properties, Applications},''
\href{http://arxiv.org/abs/1912.11100}{{\ttfamily arXiv:1912.11100 [hep-th]}}.

\bibitem{Caron-Huot:2020adz}
S.~Caron-Huot, D.~Mazac, L.~Rastelli, and D.~Simmons-Duffin, ``{Dispersive CFT
  Sum Rules},'' \href{http://arxiv.org/abs/2008.04931}{{\ttfamily
  arXiv:2008.04931 [hep-th]}}.

\bibitem{Gopakumar:2021dvg}
R.~Gopakumar, A.~Sinha, and A.~Zahed, ``{Crossing Symmetric Dispersion
  Relations for Mellin Amplitudes},''
  \href{http://arxiv.org/abs/2101.09017}{{\ttfamily arXiv:2101.09017
  [hep-th]}}.

\bibitem{Mazac:2016qev}
D.~Mazac, ``{Analytic bounds and emergence of AdS$_{2}$ physics from the
  conformal bootstrap},'' \href{http://dx.doi.org/10.1007/JHEP04(2017)146}{{\em
  JHEP} {\bfseries 04} (2017) 146},
\href{http://arxiv.org/abs/1611.10060}{{\ttfamily arXiv:1611.10060 [hep-th]}}.

\bibitem{Mazac:2018mdx}
D.~Mazac and M.~F. Paulos, ``{The Analytic Functional Bootstrap I: 1D CFTs and
  2D S-Matrices},''
\href{http://arxiv.org/abs/1803.10233}{{\ttfamily arXiv:1803.10233 [hep-th]}}.

\bibitem{Mazac:2018ycv}
D.~Mazac and M.~F. Paulos, ``{The Analytic Functional Bootstrap II: Natural
  Bases for the Crossing Equation},''
\href{http://arxiv.org/abs/1811.10646}{{\ttfamily arXiv:1811.10646 [hep-th]}}.

\bibitem{Paulos:2019gtx}
M.~F. Paulos, ``{Analytic Functional Bootstrap for CFTs in $d>1$},''
\href{http://arxiv.org/abs/1910.08563}{{\ttfamily arXiv:1910.08563 [hep-th]}}.

\bibitem{Carmi:2020ekr}
D.~Carmi, J.~Penedones, J.~A. Silva, and A.~Zhiboedov, ``{Applications of
  dispersive sum rules: $\epsilon$-expansion and holography},''
  \href{http://arxiv.org/abs/2009.13506}{{\ttfamily arXiv:2009.13506
  [hep-th]}}.

\bibitem{Giddings:2007qq}
S.~B. Giddings and M.~Srednicki, ``{High-energy gravitational scattering and
  black hole resonances},''
  \href{http://dx.doi.org/10.1103/PhysRevD.77.085025}{{\em Phys. Rev.}
  {\bfseries D77} (2008) 085025},
\href{http://arxiv.org/abs/0711.5012}{{\ttfamily arXiv:0711.5012 [hep-th]}}.

\bibitem{Correia:2020xtr}
M.~Correia, A.~Sever, and A.~Zhiboedov, ``{An Analytical Toolkit for the
  S-matrix Bootstrap},'' \href{http://arxiv.org/abs/2006.08221}{{\ttfamily
  arXiv:2006.08221 [hep-th]}}.

\bibitem{Li:2021cjv}
X.~Li, C.~Yang, H.~Xu, C.~Zhang, and S.-Y. Zhou, ``{Positivity in Multi-Field
  EFTs},'' \href{http://arxiv.org/abs/2101.01191}{{\ttfamily arXiv:2101.01191
  [hep-ph]}}.

\bibitem{Simmons-Duffin:2015qma}
D.~Simmons-Duffin, ``{A Semidefinite Program Solver for the Conformal
  Bootstrap},'' \href{http://dx.doi.org/10.1007/JHEP06(2015)174}{{\em JHEP}
  {\bfseries 06} (2015) 174}, \href{http://arxiv.org/abs/1502.02033}{{\ttfamily
  arXiv:1502.02033 [hep-th]}}.

\bibitem{Landry:2019qug}
W.~Landry and D.~Simmons-Duffin, ``{Scaling the semidefinite program solver
  SDPB},'' \href{http://arxiv.org/abs/1909.09745}{{\ttfamily arXiv:1909.09745
  [hep-th]}}.

\bibitem{Poland:2010wg}
D.~Poland and D.~Simmons-Duffin, ``{Bounds on 4D Conformal and Superconformal
  Field Theories},'' \href{http://dx.doi.org/10.1007/JHEP05(2011)017}{{\em
  JHEP} {\bfseries 05} (2011) 017},
  \href{http://arxiv.org/abs/1009.2087}{{\ttfamily arXiv:1009.2087 [hep-th]}}.

\bibitem{ElShowk:2012hu}
S.~El-Showk and M.~F. Paulos, ``{Bootstrapping Conformal Field Theories with
  the Extremal Functional Method},''
  \href{http://dx.doi.org/10.1103/PhysRevLett.111.241601}{{\em Phys. Rev.
  Lett.} {\bfseries 111} no.~24, (2013) 241601},
  \href{http://arxiv.org/abs/1211.2810}{{\ttfamily arXiv:1211.2810 [hep-th]}}.

\bibitem{Schwarz:1982jn}
J.~H. Schwarz, ``{Superstring Theory},''
  \href{http://dx.doi.org/10.1016/0370-1573(82)90087-4}{{\em Phys. Rept.}
  {\bfseries 89} (1982) 223--322}.

\bibitem{NimaTalkStrings2016}
N.~Arkani-Hamed, ``Towards deriving string theory as the weakly coupled {UV}
  completion of gravity.'' Talk at Strings 2016.

\bibitem{Kawai:1985xq}
H.~Kawai, D.~C. Lewellen, and S.~H.~H. Tye, ``{A Relation Between Tree
  Amplitudes of Closed and Open Strings},''
  \href{http://dx.doi.org/10.1016/0550-3213(86)90362-7}{{\em Nucl. Phys. B}
  {\bfseries 269} (1986) 1--23}.

\bibitem{Amati:1987wq}
D.~Amati, M.~Ciafaloni, and G.~Veneziano, ``{Superstring Collisions at
  Planckian Energies},''
  \href{http://dx.doi.org/10.1016/0370-2693(87)90346-7}{{\em Phys. Lett. B}
  {\bfseries 197} (1987) 81}.

\bibitem{Guerrieri:2021ivu}
A.~Guerrieri, J.~Penedones, and P.~Vieira, ``{Where is String Theory?},''
  \href{http://arxiv.org/abs/2102.02847}{{\ttfamily arXiv:2102.02847
  [hep-th]}}.

\bibitem{Poland:2011ey}
D.~Poland, D.~Simmons-Duffin, and A.~Vichi, ``{Carving Out the Space of 4D
  CFTs},'' \href{http://dx.doi.org/10.1007/JHEP05(2012)110}{{\em JHEP}
  {\bfseries 05} (2012) 110}, \href{http://arxiv.org/abs/1109.5176}{{\ttfamily
  arXiv:1109.5176 [hep-th]}}.

\bibitem{Kos:2014bka}
F.~Kos, D.~Poland, and D.~Simmons-Duffin, ``{Bootstrapping Mixed Correlators in
  the 3D Ising Model},'' \href{http://dx.doi.org/10.1007/JHEP11(2014)109}{{\em
  JHEP} {\bfseries 11} (2014) 109},
  \href{http://arxiv.org/abs/1406.4858}{{\ttfamily arXiv:1406.4858 [hep-th]}}.

\bibitem{NIST:DLMF}
``{\it NIST Digital Library of Mathematical Functions}.''
  Http://dlmf.nist.gov/, release 1.1.0 of 2020-12-15.
\newblock \url{http://dlmf.nist.gov/}. F.~W.~J. Olver, A.~B. {Olde Daalhuis},
  D.~W. Lozier, B.~I. Schneider, R.~F. Boisvert, C.~W. Clark, B.~R. Miller,
  B.~V. Saunders, H.~S. Cohl, and M.~A. McClain, eds.

\end{thebibliography}\endgroup

\end{document}